\documentclass[twocolumn,prb,showpacs,superscriptaddress]{revtex4-1}
\usepackage{amsfonts}
\usepackage{graphicx}
\usepackage{amsmath}
\usepackage{times}
\usepackage{amssymb}
\usepackage{color}
\usepackage[colorlinks,bookmarks=false,citecolor=blue,linkcolor=red,urlcolor=blue]{hyperref}

\newcommand{\wk}{\omega_{\vec k}}
\newcommand{\Wk}{\Omega_{\vec k}}
\newcommand{\gk}{\gamma_{\vec k}}
\newcommand{\qq}{q}     % the QPT control parameter Kd/J
\newcommand{\qqc}{q_c}  % the critical \qq
\newcommand{\lm}{\lambda} % Condensate parameter

\newcommand{\kk}{\vec k}
\newcommand{\qqi}{q_{\rm TFIM}}     % the QPT control parameter Kd/J
\newcommand{\Tb}{\tilde{T}}
\newcommand{\tb}{\tilde{\tau}}
\newcommand{\Pb}{\tilde{P}}
\newcommand{\Ab}{\tilde{A}}
\newcommand{\Bb}{\tilde{B}}
\newcommand{\ub}{\tilde{u}}
\newcommand{\vb}{\tilde{v}}
\newcommand{\Wkt}{\tilde{\Omega}_{\vec{k}}}

\newcommand{\vqad}{\Gamma_{41}^{d}}  % first tau quartic vertex
\newcommand{\vqbd}{\Gamma_{42}^{d}}  % second tau quartic vertex
\newcommand{\vqcd}{\Gamma_{43}^{d}}  % third tau quartic vertex
\newcommand{\vqdd}{\Gamma_{44}^{d}}  % fourth tau quartic vertex

\newcommand{\vqao}{\Gamma_{41}^{o}}  % first tau quartic vertex
\newcommand{\vqbo}{\Gamma_{42}^{o}}  % second tau quartic vertex
\newcommand{\vqco}{\Gamma_{43}^{o}}  % third tau quartic vertex
\newcommand{\vqdo}{\Gamma_{44}^{o}}  % fourth tau quartic vertex

\newcommand{\vcao}{\Gamma_{31}^{o}}  % first tau cubic vertex
\newcommand{\vcbo}{\Gamma_{32}^{o}}  % second tau cubic vertex
\newcommand{\ii}{\imath}   % imag unit

\newcommand{\x}{k}      % the CUT expansion parameter K/J

\begin{document}

\title{Linked-cluster expansions for quantum magnets on the hypercubic lattice}

\author{Kris Coester}
\affiliation{Lehrstuhl f\"ur Theoretische Physik I, Otto-Hahn-Str.~4, TU Dortmund, D-44221 Dortmund, Germany}
\author{Darshan G. Joshi}
\affiliation{Institut f\"ur Theoretische Physik,
Technische Universit\"at Dresden, 01062 Dresden, Germany}
\author{Matthias Vojta}
\affiliation{Institut f\"ur Theoretische Physik,
Technische Universit\"at Dresden, 01062 Dresden, Germany}
 \author{Kai Phillip Schmidt}
 \affiliation{Lehrstuhl f\"ur Theoretische Physik I, Staudtstra\ss e 7, FAU Erlangen-N\"urnberg, D-91058 Erlangen, Germany}

\date{\today}

\begin{abstract}
For arbitrary space dimension $d$ we investigate the quantum phase transitions of two paradigmatic spin models defined on a hypercubic lattice, the coupled-dimer Heisenberg model and the transverse-field Ising model. To this end high-order linked-cluster expansions for the ground-state energy and the one-particle gap are performed up to order 9 about the decoupled-dimer and high-field limits, respectively. Extrapolations of the high-order series yield the location of the quantum phase transition and the correlation-length exponent $\nu$ as a function of space dimension $d$. The results are complemented by $1/d$ expansions to next-to-leading order of observables across the phase diagrams. Remarkably, our analysis of the extrapolated linked-cluster expansion allows to extract the coefficients of the full $1/d$ expansion for the phase-boundary location in both models {\it exactly} in leading order and quantitatively for subleading corrections.
\end{abstract}

\pacs{75.40.Mg, 75.10.Jm, 75.10.Kt,  02.70.-c }

\maketitle

%
%%%%%%%%%%%%%%%%%%%%%%%%%%%%%%%%%%%%%%%%%%%%%%%%%%%%%%%%%%%%%%%%%%%%%%%%%%%%%%%%%%%%%%%
\section{Introdcution}
\label{sec:intro}
%%%%%%%%%%%%%%%%%%%%%%%%%%%%%%%%%%%%%%%%%%%%%%%%%%%%%%%%%%%%%%%%%%%%%%%%%%%%%%%%%%%%%%%
Zero-temperature phase transitions\cite{ssbook,mv_rop,ss_natph} are a central topic in many domains of modern physics, in particular in correlated-electron systems. Such a quantum phase transition (QPT), driven by varying a non-thermal parameter like magnetic field, pressure, or doping, implies a qualitative change of the system's ground state. Near a continuous QPT one observes unconventional finite-temperature behavior which is universal, i.e., independent of microscopic details.

Quantum magnets are a perfect playground, both experimentally and theoretically, to investigate quantum criticality and the associated universal behavior. Here, theoretical investigations can mainly be distinguished in two groups. The first studies continuum-limit quantum field theories using, e.g., renormalization-group techniques -- these are designed to access the long-wavelength behavior near criticality. The second works with lattice models which are analyzed using approximate analytical or numerical techniques -- such approaches capture microscopic aspects of the problem at hand, but are often unsuitable to describe critical behavior.

One powerful technique to study quantum lattice models and QPTs are linked-cluster expansions (LCEs) \cite{Oitmaa2006}. Here high-order series expansions are derived at zero temperature in the thermodynamic limit for relevant physical quantities of the microscopic model under consideration. The obtained series can then be extrapolated, giving access to quantum critical points and critical exponents. In practice, LCEs are done most efficiently via a full graph expansion, i.e.~the topologically distinct fluctuations are determined on finite graphs and are then embedded directly into the thermodynamic-limit system. LCEs therefore consist of two independent steps: the explicit calculation on graphs and the embedding procedure. The latter  is essentially the combinatorial problem to find the number of possible embeddings of a graph on the infinite lattice.

LCEs are usually performed for a microscopic lattice Hamiltonian in fixed space dimension $d$. However, since it is only the embedding procedure and the selection of graphs which depend explicitly on $d$, it is also possible to realize LCEs for arbitrary $d$ so that the high-order series of physical quantities contain $d$ a free parameter. Such LCEs for general $d$ have been set up for classical Ising and Potts models on the hypercubic lattice in Refs.~\onlinecite{Fisher90, Hellmund06}. Recently, we have extended such expansions to ground-state energies and one-particle dispersions for quantum lattice problems in Ref.~\onlinecite{Joshi2015} where they were mainly used as cross-check of the $1/d$ expansion methodology developed there.

It is the purpose of this paper to study QPTs of quantum magnets on the hypercubic lattice for arbitrary $d$ using LCEs. We have pushed the series expansions for general $d$ to considerably higher orders which allows us to investigate the QPT of both the coupled-dimer Heisenberg model (like in Ref.~\onlinecite{Joshi2015}) and the transverse-field Ising model. Specifically, we calculate high-order LCEs for the ground-state energy and for the one-particle gap about the decoupled dimer (high-field) limit for both models. Extrapolations of these quantities allow us to determine the location of the phase boundary as function of $d$ and to predict the behavior of the leading coefficients of the full $1/d$-expansion. Surprisingly, the first-order coefficient can be extracted exactly from the extrapolation.
We complement the LCE results by those from a systematic $1/d$ expansion to next-to-leading order in both the disordered and ordered phases: Such an expansion has been developed for coupled dimers in Refs.~\onlinecite{Joshi2015, Joshi2015b} and is extended here to the transverse-field Ising model.

The paper is organized as follows. In Sect.~\ref{sec:models}, we introduce the two microscopic models studied in this work. The technical
 aspects including the high-order series expansions for the ground-state energy and the one-particle gap as well as their extrapolation are contained in Sect.~\ref{sec:pcut}, while details of the $1/d$ expansions are given in Sect.~\ref{sec:d}.
 The physical implications and findings are discussed in Sect.~\ref{sec:results} and we conclude our work in Sect.~\ref{sec:conclusion}.
%
%%%%%%%%%%%%%%%%%%%%%%%%%%%%%%%%%%%%%%%%%%%%%%%%%%%%%%%%%%%%%%%%%%%%%%%%%%%%%%%%%%%%%%%
\section{Models}
\label{sec:models}
%%%%%%%%%%%%%%%%%%%%%%%%%%%%%%%%%%%%%%%%%%%%%%%%%%%%%%%%%%%%%%%%%%%%%%%%%%%%%%%%%%%%%%%
We consider two archetypical models on the hypercubic lattice with dimension $d$: i) a coupled-dimer Heisenberg magnet (CDHM) of spins $1/2$ and ii) the transverse-field Ising model (TFIM). In both cases an infinite-$d$ limit exists which features a non-trivial QPT.

%%%%%%%%%%%%%%%%%%%%%%%%%%%%%%%%%%%%%%%%%
\subsection{Coupled-dimer Heisenberg model}
\label{ssec:cdhm}
%%%%%%%%%%%%%%%%%%%%%%%%%%%%%%%%%%%%%%%%%
The CDHM is a model of dimers, i.e., pairs of spins 1/2, on the sites $i$ of a hypercubic lattice. The Hamiltonian reads
\begin{equation}
\label{hh}
\mathcal{H}_{\rm CDHM} =
J \sum_i \vec{S}_{i1} \cdot \vec{S}_{i2} +
\sum_{\langle ij\rangle} (K^{11} \vec{S}_{i1} \cdot \vec{S}_{j1} + K^{22} \vec{S}_{i2} \cdot \vec{S}_{j2})\,,
\end{equation}
where $\sum_{\langle ij\rangle}$ denotes a summation over pairs of nearest-neighbor dimer
sites $i,j$, and $1,2$ refer to the individual spins on each dimer. We restrict our attention to the symmetric case of $K^{11}=K^{22}\equiv K$.
For $d=1$ and $d=2$ the spin lattice of $\mathcal{H}_{\rm CDHM}$ in Eq.~\eqref{hh} corresponds to the much-studied two-leg ladder and square-lattice bilayer magnets, respectively.

A non-trivial limit $d\to\infty$ is obtained if the inter-dimer coupling constant $K$ is scaled as $1/d$ in order to preserve a competition between the $K$ and $J$ terms in \eqref{hh}. For $d\geq 2$ and $K,J>0$, the dimensionless parameter
\begin{equation}
\qq_{\rm CDHM} = \frac{Kd}{J}
\end{equation}
controls a QPT between a singlet paramagnet with gapped triplon excitations \cite{schmi03} at small $\qq$ and a gapless antiferromagnet with ordering wavevector $\vec{Q}=(\pi,\pi,\ldots)$ at large $\qq$. For $d=2$ this transition occurs at\cite{sandvik06} $\qqc = 0.793$ while $\qqc = 0.620$ has been found recently for $d=3$ in Ref.~\onlinecite{Qin2015}.

The upper critical dimension of the QPT, with O(3) order parameter, is $d=3$. Consequently, the critical exponents take mean-field values $\alpha=0$, $\beta=1/2$, $\gamma=1$, $\delta=3$, and $\nu=1/2$ for all dimensions $d\geq3$.
In contrast, one expects continuously varying anomalous exponents for $1<d<3$. For $d=3$ one finds multiplicative logarithmic corrections to mean-field behavior \cite{Kenna1993,Kenna1994,Qin2015}.

%%%%%%%%%%%%%%%%%%%%%%%%%%%%%%%%%%%%%%%%%
\subsection{Transverse-field Ising model}
\label{ssec:tfim}
%%%%%%%%%%%%%%%%%%%%%%%%%%%%%%%%%%%%%%%%%
The TFIM is described by the Hamiltonian
\begin{equation}
\label{tfim}
\mathcal{H}_{\rm TFIM} = -h\sum_i \sigma^{x}_i -J \sum_{\langle i j\rangle} \sigma^{z}_i\sigma^{z}_j\,,
\end{equation}
where $\sum_{\langle ij\rangle}$ denotes a summation over nearest-neighbor spins $i,j$ on the hypercubic lattice and $\sigma^\alpha_i$ with $\alpha\in\{x,y,z\}$ are Pauli matrices representing spin-1/2 degrees of freedom. In the following we assume a ferromagnetic Ising exchange $J>0$, but our results are equally valid for the antiferromagnetic case, since both cases can be mapped onto each other via a sublattice rotation on the hypercubic lattice.

A non-trivial limit $d\to\infty$ is obtained again by scaling the Ising coupling constant $J$ with $1/d$. For $d\geq 1$, the dimensionless parameter
\begin{equation}
\label{eq:q_tfim}
\qq_{\rm TFIM} = \frac{Jd}{h}
\end{equation}
controls the QPT between two gapped phases, the field-polarized phase at small $\qq$ and the ordered ferromagnetic state with broken $\mathbb{Z}_2$ symmetry at large $\qq$. For $d=1$, this model reduces to the well-known Ising chain in a transverse field, with the transition located at $\qqc = 1$. For $d=2$, the quantum critical point has been determined by series expansions and quantum Monte Carlo and is known to be\cite{He1990,Oitmaa1991,Bloete1995,deng_qmc3d,wessel_tfim} \mbox{$\qqc \approx 0.657$}. For the simple cubic lattice ($d=3$) series expansions and quantum Monte Carlo finds\cite{Zheng1994,deng_qmc3d} $\qqc \approx 0.582$.

The order-parameter symmetry of the TFIM is $\mathbb{Z}_2$. The upper critical dimension is $d=3$ as well, implying mean-field exponents for $d\geq3$, continuously varying exponents for $1\leq d<3$ (different from the ones of the CDHM), and multiplicative logarithmic corrections \cite{Larkin1969,Brezin1973,Wegner1973,Zheng1994} for $d=3$.

\section{Linked-cluster expansion for arbitrary $d$}
\label{sec:pcut}

In this section we provide the relevant technical aspects with respect to linked-cluster expansions for arbitrary dimension $d$ for the disordered phases of the CDHM and the TFIM on the hypercubic lattice. In practice, we calculated high-order series expansions for the ground-state energy per dimer (spin) and for the one-particle gap up to order $9$ in the relative strength of the inter-dimer (inter-site) coupling $\x = K/J$ ($\x = J/h$) for the CDHM (TFIM).

\subsection{Method}

We start by sketching the underlying method of the expansion; for details we refer the reader to Refs.~\onlinecite{gsu00,Knetter03_1,Coester2015}.
The expansion's reference point corresponds to $\x=0$. Here the ground state is given by a product state in both models. For the CDHM there are singlets on the dimers, and elementary excitations are local triplets with excitation energy $\Delta^{\rm CDHM}/J=1$. In contrast, the TFIM is in the fully polarized state with all spins pointing along $\hat x$, and the elementary excitations are local spin flips corresponding to a spin along $-\hat x$ with excitation energy $\Delta^{\rm TFIM}/2h=1$.

After a global energy shift, we can rewrite both models in the form
\begin{eqnarray}
\label{h_pert}
\mathcal{H}&=&\mathcal{H}_0+k\, \hat{V} \quad ,
\end{eqnarray}
where $\mathcal{H}_0$ has an equidistant spectrum bounded from below counting the number of excitations, namely triplets (spin flips) for the CDHM (TFIM). The perturbing parts can then be written as
\begin{align}
\hat{V}=\hat{T}_{-2}+\hat{T}_0+\hat{T}_{2} \, ,
\end{align}
where $\hat{T}_m$ changes the total number of excitations by \mbox{$m\in\{\pm 2, 0\}$}. Note that only terms with even $m$  appear in $\hat{V}$ due to the exact reflection symmetry of both models leading to a conserved parity quantum number.

Each operator $\hat{T}_m$ is a sum over local operators connecting two nearest-neighbor sites (either dimers or spins). One can therefore write
\begin{align}
\hat{T}_m=\sum_l \hat{\tau}_{m,l}\,,\label{tausumme}
\end{align}
with $\hat{\tau}_{m,l}$ effecting only the two sites connected by the link $l$ on the lattice.

The perturbative continuous unitary transformations (pCUTs) \cite{gsu00,Knetter03_1,Coester2015} map the original Hamiltonian to an effective quasiparticle conserving Hamiltonian of the form
\begin{eqnarray}
\hat{H}_\text{eff}(k)=\hat{H}_0+\sum_{n=1}^{\infty}\x^n \hspace*{-2mm}
\sum_{{\rm dim}(\underline{m})=n \atop  \,M(\underline{m})=0} \hspace*{-2mm}
C(\underline{m})\,\hat{T}_{m_1}\dots \hat{T}_{m_n},\label{H_eff}
\end{eqnarray}
where $n$ reflects the perturbative order. The second sum is taken over all possible vectors $\underline{m}\!\equiv\!(m_1, \ldots, m_n)$ with $m_i\in\{\pm 2, 0\}$ and dimension ${\rm dim}(\underline{m})=n$. Each term of this sum is weighted by the rational coefficient \mbox{$C(\underline{m})\in \mathbb{Q}$} which has been calculated model-independently up to high orders.\cite{gsu00} The additional restriction $M(\underline{m})\equiv\sum m_i=0$ reflects the quasiparticle-conserving property of the effective Hamiltonian, i.e.~, the resulting Hamiltonian is block-diagonal in the number of quasiparticles $[\hat{H}_\text{eff},\hat{H}_0]=0$. Each quasiparticle block can then be investigated separately which represents a major simplification of the complicated many-body problem.

The operator products $\hat{T}_{m_1}\dots \hat{T}_{m_n}$ appearing in order $n$ can be interpreted as virtual fluctuations of ``length'' $l\leq n$ leading to dressed quasiparticles. According to the linked-cluster theorem, only linked fluctuations can have an overall contribution to the effective Hamiltonian $\hat{H}_\text{eff}$. Hence, the properties of interest can be calculated in the thermodynamic limit by applying the effective Hamiltonian on finite clusters.

Considering all linked fluctuations on the lattice (for arbitrary $d$), it becomes clear that the contribution of each fluctuation only depends on its topology. The calculation separates therefore into two independent steps: (i) The model-dependent part of the calculation is performed on the finite set of topologically distinct graphs. (ii) The model-independent part corresponds to the combinatorial problem of how many times the contribution on the graphs has to be embedded into the lattice in order to extract thermodynamic-limit properties. For the calculation in $d$ dimensions, it is the second model-independent part which represents the real challenge. As a consequence, we reach the same maximum order 9 for both models, both for the ground-state energy and the one-particle gap.

%%%%%%%%%%%%%%%%%%%%%%%%%%%%%%%%%%%%%%%%%%%%%%%%%%%%%%%%%%%%%%%%%%%%%%%%%%%%%%%%%%%%%%%%%%%%%%%%%%%%%%%%%%%%%%%%%%%%%%%
\subsection{Ground-state energy}
%%%%%%%%%%%%%%%%%%%%%%%%%%%%%%%%%%%%%%%%%%%%%%%%%%%%%%%%%%%%%%%%%%%%%%%%%%%%%%%%%%%%%%%%%%%%%%%%%%%%%%%%%%%%%%%%%%%%%%%

In the following, we discuss the calculation of the ground-state energy per dimer (spin) $\epsilon_0^{\rm CDHM}$ ($\epsilon_0^{\rm TFIM}$) for the hypercubic CDHM (TFIM) for arbitrary $d$. The calculation is performed up to order $\x^9$, using pCUTs and a full graph decomposition. The first step of the calculation is conventional and it is part of any LCE. One determines the (reduced) contributions to the ground-state energy $\epsilon_{0,n}$, which is done in pCUTs by simply applying Eq.~\eqref{H_eff} to the zero-particle state on the relevant graphs. In order to avoid double counting of contributions, the reduced contribution $\epsilon_{0,i}$ to $\epsilon_0$ of each graph $\mathcal{G}_i$ has to be calculated by subtracting the contributions of all subgraphs.

Up to perturbative order $n$, only graphs up to $n$ links have to be considered due to the linked-cluster theorem. Additionally, it is useful to check whether the graphs fit onto the lattice and whether each graph has a finite contribution in the order under consideration. The latter depends on both, the model and the observable. In the case of the ground-state energy one has a specific selection rule for both models that each link has to be touched twice by the perturbation as long as it is not part of a closed loop of links \cite{Joshi2015}. This property drastically reduces the total number of graphs which one has to treat.

The embedding factor $\nu_i(d)$ for graph $\mathcal{G}_i$, being the number of possible embeddings of $\mathcal{G}_i$ on the lattice, is a function of the spatial dimension $d$. The ground-state energy in the thermodynamic limit is then given by
\begin{align}
\epsilon_0 =\sum_i \nu_i(d)\, \epsilon_{0,i}\,.
\end{align}
The determination of the embedding factors $\nu_i (d)$ for arbitrary $d$ is the most challenging part of the calculation. Note that the embedding factors $\nu_i (d)$ are exactly the same for both models; only the contributions $\epsilon_{0,i}$ are model-dependent. To determine the embedding factor in $d$ dimensions, we apply a scheme similar to the one presented in Ref.~\onlinecite{Hellmund06}. Each graph can be associated with a dimension defined by the number of dimensions that the graph can maximally occupy. The $d$-dependent embedding factor is then given by a polynomial in $d$ of degree $m_\text{max}$ which is determined by the embedding factors of the graph in dimension $d=1$, $d=2, \dots, d=m_\text{max}$.

In order to determine the embedding factors it is necessary to divide the number of naive embeddings by the symmetry factor $S_i$ of $\mathcal{G}_i$. Otherwise one overcounts contributions, since embeddings connected by a symmetry mapping of the graph represent exactly the same fluctuation on the lattice in the thermodynamic limit.

Following these principles, we find the following ground-state energy per dimer for the CDHM
\begin{align}
\label{e0_smallx_cd}
\frac{\epsilon_0^{\rm CDHM}}{J} &= -\frac{3}{4}
-\frac{3}{8}d\,{\x}^{2}-\frac{3}{16}d\,{\x}^{3}
+\Big(\frac{21}{128}d-\frac{9}{64}d^2\Big){\x}^{4}\notag\\
&+\Big(\frac {57}{256}d-\frac{3}{64}d^2\Big){\x}^{5}
+\Big(-\frac {2781}{1024}d + \frac {273}{64}d^2  \notag\\
&-\frac{357}{256}d^3 \Big){\x}^{6} +\Big(-\frac {73293}{16384}d  + \frac {53205}{8192}d^2  \notag\\
&- \frac{8499}{4096}d^3  \Big){\x}^{7}+\Big( \frac{1151577}{32768} d-\frac{2270385}{32768} d^2\notag\\
&+\frac{687885}{16384} d^3-\frac{8313}{1024} d^4\Big) {\x}^{8} +\Big( \frac{80239263}{1048576} d\notag\\
&-\frac{75882381}{524288} d^2+\frac{21745125}{262144} d^3-\frac{490731}{32768} d^4 \Big) {\x}^{9}
\end{align}
and the ground-state energy per spin for the TFIM
\begin{align}
\label{e0_smallx_tf}
\frac{\epsilon_0^{\rm TFIM}}{2h} &= -\frac{1}{2} -\frac{1}{8} {\x}^2d+\Big(\frac{13}{128}d-\frac{7}{64} d^2\Big) {\x}^{4} \\
& +\Big( -\frac{367}{512}d+\frac{311}{256}d^2-\frac{1}{2}d^3 \Big) {\x}^{6} \notag\\
& +\Big(\frac{7031}{512}d-\frac{947263}{32768}d^2+\frac{321187}{16384}d^3-\frac{4535}{1024}d^4 \Big) {\x}^{8}\notag\,.
\end{align}

One can easily check that these results match the ones from literature for specific dimensions. For $d=1$, this formula reduces to the known results for the CDHM on a two-leg ladder \cite{Knetter03_2} and for the TFIM on a chain \cite{Pfeuty1970,Oitmaa2006}. For $d=2$, we reproduce the numerical results of the ground-state energy for the square-lattice Heisenberg bilayer \cite{Weihong97} and the square lattice TFIM \cite{He1990}. For $d=3$, the series expansion for the TFIM is known \cite{Zheng1994} while for the CDHM the high-order series expansion was unknown.

\subsection{One-particle gap}

For the one-particle dispersion $\wk$, the embedding procedure is more demanding, since each embedding is associated with different hopping elements in the thermodynamic limit. However, if one is only interested in the one-particle gap which is located at $\vec k=\vec Q$ for the CDHM and at $\vec k=0$ for the TFIM, it is possible to apply a similar scheme as for the ground-state energy:

The hypercubic lattice is bipartite, i.e., it can be divided into two sublattices. The hopping elements can then be classified into hopping elements on the same sublattice and hopping elements between both sublattices. At $\vec k=0$, relevant for the TFIM, the contribution of all hopping elements is simply given by the embedding factor. However, at $\vec k=(\pi,\pi,\ldots,\pi)$ relevant for the CDHM, only hopping elements on the same sublattice are given by the embedding factor, while the contribution of hopping elements between different sublattices must be modified with an additional sign due to the gap momentum. Consequently, the one-particle gap $\Delta$ of the CDHM is given by
\begin{equation}
\Delta=\sum_{i} \nu_i (d) \Delta^{\mathcal{G}_i}
\end{equation}
with
\begin{equation}
\Delta^{\mathcal{G}_i}=\sum_{l,m} \sigma_{l,m} t^{(i)}_{l,m}
\end{equation}
where the double sum runs over all sites of the graph $\mathcal{G}_i$, $t^{(i)}_{l,m}$ corresponds to the (reduced) one-particle hopping element from site $l$ to site $m$ determined by the pCUT calculation, and $\sigma_{l,m}=1$ ($\sigma_{l,m}=-1$) if the number of links between sites $l$ and $m$ is even (odd). The latter represents the additional sign to account for the momentum $\vec Q$ of the one-particle gap in the case of the CDHM.

The resulting one-particle gaps for both models are given by
\begin{align}
\label{gap_smallx_cd}
\frac{\Delta^{\rm CDHM}}{J} &= 1-d{\x}+\Big( d-\frac{1}{2}d^2\Big){\x}^{2} +\Big( \frac{1}{4}d+\frac{1}{2}d^2 \notag\\
 & -\frac{1}{2}d^3\Big){\x}^{3} +\Big( -\frac{5}{4}d+\frac{3}{4}d^2+d^3-\frac{5}{8}d^4 \Big){\x}^{4} \notag\\
 & +\Big( \frac{29}{128}d-\frac{5}{2}d^2+\frac{9}{8}d^3  +\frac{7}{4}d^4-\frac{7}{8}d^5\Big){\x}^{5} \notag\\
 & +\Big( \frac{17169}{1024}d-\frac{12585}{512}d^2+\frac{1429}{256}d^3+\frac{17}{64}d^4\notag\\
 & +\frac{25}{8}d^5-\frac{21}{16}d^6\Big){\x}^{6} +\Big( -\frac{1845}{4096}d+\frac{94275}{4096}d^2\notag\\
 & -\frac{29833}{1024}d^3+\frac{4117}{1024}d^4-\frac{213}{256}d^5+\frac{91}{16}d^6 \notag\\
 & -\frac{33}{16}d^7 \Big){\x}^{7}+\Big( -\frac{31976937}{65536}d+\frac{30652045}{32768}d^2\notag\\
 & -\frac{133669}{256}d^3+\frac{580521}{8192}d^4+\frac{553}{512}d^5-\frac{4059}{1024}d^6\notag\\
 & +\frac{21}{2}d^7-\frac{429}{128}d^8 \Big){\x}^{8} +\Big( -\frac{128426725}{524288}d\notag\\
 & -\frac{36029001}{262144}d^2+\frac{110976899}{131072}d^3-\frac{34605481}{65536}d^4\notag\\
 & +\frac{491585}{8192}d^5+\frac{11615}{8192}d^6-\frac{47333}{4096}d^7\notag\\
 & +\frac{627}{32}d^8-\frac{715}{128}d^9\Big){\x}^{9} \,.
\end{align}
for the CDHM and
\begin{align}
\label{gap_smallx_tf}
\frac{\Delta^{\rm TFIM}}{2h} &=  1-dx+\Big( \frac{1}{2}d -\frac{1}{2}d^2 \Big) {\x}^{2} +\Big( -\frac{1}{4}d+\frac{3}{4}d^2 \notag\\
& -\frac{1}{2}d^3 \Big) {\x}^{3} +\Big( -\frac{5}{8}d+\frac{1}{4}d^2+d^3-\frac{5}{8}d^4 \Big) {\x}^{4} \notag\\
& +\Big( \frac{5}{4}d-\frac{5}{2}d^2+\frac{1}{2}d^3+\frac{13}{8}d^4-\frac{7}{8}d^5 \Big) {\x}^{5} \notag\\
& +\Big( \frac{3873}{512}d-\frac{5577}{512}d^2+\frac{273}{128}d^3-\frac{39}{128}d^4+\frac{45}{16}d^5 \notag\\
& -\frac{21}{16}d^6 \Big) {\x}^{6}  +\Big( -\frac{9635}{512}d+\frac{95253}{2048}d^2-\frac{72089}{2048}d^3 \notag\\
& +\frac{3165}{512}d^4-\frac{841}{512}d^5+\frac{161}{32}d^6-\frac{33}{16}d^7 \Big) {\x}^{7} \notag\\
& +\Big( -\frac{7102997}{32768}d+\frac{13840105}{32768}d^2-\frac{970563}{4096}d^3 \notag\\
& +\frac{205211}{8192}d^4+\frac{21305}{4096}d^5-\frac{9695}{2048}d^6 +\frac{147}{16}d^7 \notag\\
& -\frac{429}{128}d^8 \Big) {\x}^{8} +\Big( \frac{67215987}{131072}d-\frac{12296139}{8192}d^2 \notag\\
& +\frac{205457057}{131072}d^3  -\frac{5394851}{8192}d^4+\frac{2336733}{32768}d^5 \notag\\
& +\frac{134797}{16384}d^6-\frac{96289}{8192}d^7+\frac{1089}{64}d^8-\frac{715}{128}d^9 \Big) {\x}^{9}
\end{align}
for the TFIM. As for the ground-state energy, these results match with the ones known from literature\cite{Knetter03_2,Oitmaa2006,Weihong97,He1990,Zheng1994} for specific values of $d$.

%%%%%%%%%%%%%%%%%%%%%%%%%%%%%%%%%%%%%%%%%%%%%%%%%%%%%%%%%%%%%%%%%%%%%%%%%%%%%%%%%%%%%%%%%%%%%%%%%%%%%%%%%%%%%%%%%%%%%%%
\subsection{Extrapolation}
\label{ssec:extrapol}

The obtained LCEs for the physical quantities have to be extrapolated in order to locate
quantum critical points and to determine the associated critical exponents. For a general review on series extrapolation we refer to Ref.~\onlinecite{Guttmann1989}. Here we give the relevant information on the specific extrapolation techniques we applied.

One expects the following critical behavior close to a QPT:
\begin{eqnarray}
 \label{eq:critical}
 \frac{\partial^2 e_0}{\partial\x^2} &\propto & \left( \x-\x_{\rm c}\right)^{-\alpha} \nonumber\\
 \Delta &\propto & \left( \x-\x_{\rm c}\right)^{z\nu}\quad ,
\end{eqnarray}
where $\alpha$, $z$, and $\nu$ are the specific-heat, dynamic, and correlation length exponents, respectively. The models under consideration have $z=1$.

Our series are all of the form
\begin{align}
F(\x)=\sum_{n\geq 0}^m a_n \x^n=a_0+a_1\x+a_2\x^2+\dots a_m\x^m,
\end{align}
with $\x\in \mathbb{R}$ and $a_i \in \mathbb{R}$. If one has power-law behavior near a critical value $\x_{\rm c}$ like in Eqs.~\eqref{eq:critical}, the true physical function $\tilde{F}(\x)$ close to $\x_{\rm c}$ is given by
\begin{align}
\tilde{F}(\x)\approx \left(1-\frac{\x}{\x_{\rm c}}\right)^{-\theta} A(\x),
\end{align}
where $\theta$ is the associated critical exponent. If $A(\x)$ is analytic at $\x=\x_{\rm c}$, we can write
\begin{align}
\label{eq:Ftilde}
\tilde{F}(\x)\approx \left(1-\frac{\x}{\x_{\rm c}}\right)^{-\theta}A|_{\x=\x_{\rm c}}\left(1+\mathcal{O}\left(1-\frac{\x}{\x_{\rm c}}\right)\right).
\end{align}
Near the critical value $\x_{\rm c}$, the logarithmic derivative is then given by
\begin{align}
\tilde{D}(\x)&:=\frac{\text{d}}{\text{d}\x}\ln{\tilde{F}(\x)}\label{dx}\\
&\approx \frac{\theta}{\x_{\rm c}-\x}\left\{ 1+ \mathcal{O}(\x-\x_{\rm c})\right\}\nonumber.
\end{align}
In the case of power-law behavior, the logarithmic derivative $\tilde{D}(\x)$ is therefore expected to exhibit a single pole.

The latter is the reason why so-called Dlog-Pad\'{e} extrapolation is often used to extract critical points and critical exponents from high-order series expansions. Dlog-Pad\'e extrapolants of $F(\x)$ are defined by
\begin{align}
\label{eq:dlogP1}
dP[L/M]_F(\x)=\exp\left(\int_{0}^\x P[L/M]_{D}\,\,\text{d}\x'\right)
\end{align}
and represent physically grounded extrapolants in the case of a second-order phase transition. Here $P[L/M]_{D}$ denotes a standard Pad\'e extrapolation of the logarithmic derivative
\begin{align}
\label{eq:dlogP2}
P[L/M]_{D}:=\frac{P_L(\x)}{Q_M(\x)}=\frac{p_0+p_1\x+\dots + p_L \x^L}{q_0+q_1\x+\dots q_M \x^M}\quad,
\end{align}
with $p_i\in \mathbb{R}$ and $q_i \in \mathbb{R}$ and $q_0=1$. Additionally, $L$ and $M$ have to be chosen so that $L+M-1\leq m$. Physical poles of $P[L/M]_{D}(\x)$ then indicate critical values $\x_{\rm c}$ while the corresponding critical exponent of the pole $\x_{\rm c}$ can be deduced by
\begin{align}
\theta\equiv\left.\frac{P_L(\x)}{\frac{\text{d}}{\text{d}\x} Q_M(\x)}\right|_{\x=\x_{\rm c}} \label{extract_exponent}.
\end{align}
If the exact value (or a quantitative estimate from other approaches) of $\x_{\rm c}$ is known, one can obtain better estimates of the critical exponent by defining
\begin{align*}
\theta^*(\x)&\equiv(\x_{\rm c}-\x)D(\x)\\
&\approx \theta+\mathcal{O}(\x-\x_{\rm c}),
\end{align*}
where $D(\x)$ is given by Eq.(\ref{dx}). Then
\begin{align}
P[L/M]_{\theta^*}\big|_{\x=\x_{\rm c}}=\theta \label{biasnue}
\end{align}
yields a (biased) estimate of the critical exponent.

At the upper critical dimension $d=3$, both models display multiplicative corrections to
 Eqs.~\eqref{eq:critical} close to the quantum critical point so that one expects the following
 critical behavivour
\begin{align}
\bar{F}(\x)\approx \left(1-\frac{\x}{\x_{\rm c}}\right)^{-\theta} \left(\ln\left( 1-\frac{\x}{\x_{\rm c}}\right)\right)^{p} \bar{A}(\x),
\end{align}
where $\x_{\rm c}$ ($\theta$) are the associated critical point (exponent) as before while $p$ yields the power of multiplicative logarithmic corrections. Clearly, the extraction of $p$ from a high-order series expansion is very demanding. The only reasonable approach is to bias the extrapolation by fixing $\x_{\rm c}$ and $\theta$, e.g., the critical exponents $\theta$ are given by the well-known mean-field values.

Assuming again that the function $\bar{A}(\x)$ is analytic close to $\x_{\rm c}$, Eq.~\eqref{eq:Ftilde} transforms into
\begin{eqnarray}
\label{eq:Fbar}
\bar{F}(\x)&\approx& \left(1-\frac{\x}{\x_{\rm c}}\right)^{-\theta} \left(\ln\left( 1-\frac{\x}{\x_{\rm c}}\right)\right)^{p}\bar{A}|_{\x=\x_{\rm c}}\nonumber\\
           && \cdot\left(1+\mathcal{O}\left(1-\frac{\x}{\x_{\rm c}}\right)\right).
\end{eqnarray}
and the logarithmic derivative Eq.~\eqref{dx} becomes
\begin{align}
\bar{D}(\x)&\approx \frac{\theta}{\x_{\rm c}-\x} + \frac{-p}{\ln\left(1-\x/\x_{\rm c}\right)\left(\x_{\rm c}-\x\right)} + \mathcal{O}\left(\x-\x_{\rm c}\right)\nonumber.
\end{align}
One can then estimate the multiplicative logarithmic correction $p$ by defining
\begin{align*}
  p^{*}(\x)&\equiv -\ln\left( 1-\x/\x_{\rm c}\right) \left[  \left( \x_{\rm c}-\x\right) D(\x)-\theta \right]\\
           &\approx p +\mathcal{O}(\x-\x_{\rm c}),
\end{align*}
and by performing Pad\'{e} extrapolants of this function
\begin{align}
P[L/M]_{p*}\big|_{\x=\x_{\rm c}}=p \label{biasp}\quad .
\end{align}
%
%%%%%%%%%%%%%%%%%%%%%%%%%%%%%%%%%%%%%%%%%%%%%%%%%%%%%%%%%%%%%%%%%%%%%%%%%%%%%%%%%%%%%%%
%%%%%%%%%%%%%%%%%%%%%%%%%%%%%%%%%%%%%%%%%%%%%%%%%%%%%%%%%%%%%%%%%%%%%%%%%%%%%%%%%%%%%%%
\section{$1/d$ expansion}
\label{sec:d}
%%%%%%%%%%%%%%%%%%%%%%%%%%%%%%%%%%%%%%%%%%%%%%%%%%%%%%%%%%%%%%%%%%%%%%%%%%%%%%%%%%%%%%%

Utilizing the limit of large spatial dimension $d$, an analytic $1/d$ expansion for spin models with order--disorder QPT was developed in Refs.~\onlinecite{Joshi2015} and \onlinecite{Joshi2015b}. In the limit $d \to \infty$, non-local fluctuations are suppressed, such that a suitable product-state wavefunction can be used as a reference state. The $1/d$ expansion is obtained from a theory of interacting bosons which capture fluctuations on top of the product state; in this theory, factors of $1/d$ do not appear in the Hamiltonian but are {\em generated} via momentum summations. For large $d$, critical exponents take mean-field values which allows one to identify observables which are analytic even at the quantum critical point. This paves the way to a fully analytic description across the QPT.

Refs.~\onlinecite{Joshi2015} and \onlinecite{Joshi2015b} developed and applied this methodology to a model of coupled dimers on the hypercubic lattice. In the present paper, we will make use of those results in Sect.~\ref{sec:results} below. In addition, we apply and extend the $1/d$ expansion method to the transverse-field Ising model via a suitable auxiliary-boson description. In this section, $\qq \equiv \qqi$ is the tuning parameter defined in Eq.~\eqref{eq:q_tfim}.

%%%%%%%%%%%%%%%%%%%%%%%%%%%%%%%%%%%%%%%%%%
\subsection{Quantum paramagnetic phase}
\label{sec:qp}
%%%%%%%%%%%%%%%%%%%%%%%%%%%%%%%%%%%%%%%%%%

Let us first discuss $1/d$ expansion in the quantum paramagnetic phase of the TFIM, where a suitable reference state is given by
\begin{equation}
\label{eq:qp_ref}
\Psi = \prod_{i} | 0 \rangle_{i} \,,~~~| 0 \rangle_{i} = \frac{| \uparrow \rangle_{i} + | \downarrow \rangle_{i}}{\sqrt{2}},
\end{equation}
%In the above expression $i$ denotes the lattice site and
%\begin{equation}
%\label{eq:sx_eig1}
%| 0 \rangle_{i} = \frac{| \uparrow \rangle_{i} + | \downarrow \rangle_{i}}{\sqrt{2}}
%\end{equation}
where $i$ denotes lattice site and $|0\rangle_i$ is an eigenstate of $\sigma^{x}_{i}$ with an eigenvalue $1$. A local excitation is given by
\begin{equation}
\label{eq:sx_eig2}
| T \rangle_{i} = \frac{| \uparrow \rangle_{i} - | \downarrow \rangle_{i}}{\sqrt{2}}  \,,
\end{equation}
which is an eigenstate of $\sigma^{x}_{i}$ with an eigenvalue $-1$. We can introduce an auxiliary boson to efficiently describe this process as follows:
\begin{equation}
\label{eq:Tdef}
T_{i}^{\dagger} | 0 \rangle_i = | T \rangle_i \,;~~~~~~ T_{i} | 0 \rangle_{i} = 0 \,,
\end{equation}
such that it satisfies the usual bosonic commutation relations.
%\begin{equation}
%\label{eq:Tcomm}
%\big[ T_{i}, T_{j}^{\dagger} \big] = \delta_{i,j} \,;~
%\big[ T_{i}, T_{j} \big] = \big[ T_{i}^{\dagger}, T_{j}^{\dagger} \big] = 0 \,.
%\end{equation}
%It is straightforward to evaluate the matrix elements of the spin operators w.r.t. the eigenstates of $S^{x}$ to obtain
%\begin{align}
%\label{eq:sx1}
%S^{x} &= \frac{1}{2} \left( |0\rangle\langle 0| - |T\rangle\langle T | \right) \,, \\
%\label{eq:sy1}
%S^{y} &= \frac{\ii}{2} \left( |0\rangle\langle T| - |T\rangle\langle 0 | \right) \,, \\
%\label{eq:sz1}
%S^{z} &= \frac{1}{2} \left( |0\rangle\langle T| + |T\rangle\langle 0 | \right) \,.
%\end{align}
The physical Hilbert space on each lattice site consists of only two states. This implies a hard-core constraint for the $T_i$ bosons which we implement by introducing a projection operator\cite{collins1}
\begin{equation}
\label{eq:proj_qp}
P_i = 1 - T_{i}^{\dagger} T_{i}^{\phantom{\dagger}}
\end{equation}
which is then used to express the spin operators in terms of $T_i$ bosons:
\begin{align}
\label{eq:sxdef_qp}
\sigma_{i}^{x} &=  1 - 2 T_{i}^{\dagger} T_{i}^{\phantom{\dagger}}  \,, \\
\label{eq:sydef_qp}
\sigma_{i}^{y} &= \ii \left( P_{i} T_{i} - T_{i}^{\dagger} P_{i}^{\phantom{\dagger}} \right) \,, \\
\label{eq:szdef_qp}
\sigma_{i}^{z} &= P_{i} T_{i} + T_{i}^{\dagger} P_{i}^{\phantom{\dagger}}  \,.
\end{align}
Note that the usual spin commutation relations are satisfied within the physical Hilbert space.

Now, substituting Eqs.~\eqref{eq:sxdef_qp} - \eqref{eq:szdef_qp} in the spin Hamiltonian \eqref{tfim} we obtain an interacting boson Hamiltonian (see Appendix \ref{app:large_d_qp}). We use diagrammatic perturbation theory to treat the interacting boson piece in the Hamiltonian. In the large-$d$ formalism, corrections to observables due to these terms are suppressed in powers of $1/d$, thereby leading to a systematic $1/d$ expansion. This follows from the fact that due to particular  momentum-summation properties of the Fourier-transformed interaction, all self-energy diagrams can be expressed as a series in $1/d$ in the limit $d \to \infty$. The diagrammatic treatment in this case is similar to the one discussed in Ref. \onlinecite{Joshi2015}. We will therefore only quote the $1/d$ expansion of relevant observables here while relegating technical details and relevant expressions to Appendix \ref{app:large_d_qp}.

Let us begin with the mode dispersion $\Wk$ corresponding to the single-particle excitation. Since the critical exponents $\nu$ and $z$ have mean-field values $1/2$ and $1$ respectively, the square of the excitation-energy gap is analytic at the critical point. We therefore have an analytic $1/d$ expansion for the square of the mode dispersion as follows:
\begin{equation}
\label{eq:isd_disp}
\frac{\Wk^2}{4 h^2} = 1 - 2 \gk \qq + \frac{\qq^2 }{4 d} (4 + 2 \gk \qq) \,.
\end{equation}
Here,
\begin{equation}
\label{eq:gamma_def}
\gk = \frac{1}{d} \sum_{i=1}^{d} \cos k_{i}
\end{equation}
is the interaction structure factor such that $\gk \in \left[ -1, 1 \right]$. The leading part i.e.~$\mathcal{O}(1/d^{0})$ result in Eq.~\eqref{eq:isd_disp} arises from the non-interacting boson piece ({\em harmonic approximation}) in the Hamiltonian, while contribution from the boson-interaction terms start at $\mathcal{O}(1/d)$. Despite its appearance to this order, this is {\em not} an expansion in $\qq$. As discussed at length in Ref.~\onlinecite{Joshi2015}, one can as well convert Eq.~\eqref{eq:isd_disp} into a $1/d$ expansion for $\Wk$, but this will not be well-defined at the QPT for $\vec{k}=\vec{Q}=0$.

The minimum of the mode dispersion \eqref{eq:isd_disp} is at $\vec{Q}$. Thus the excitation energy gap is given by $\Delta = \Omega_{\vec{Q}}$. Substituting $\gamma_{\vec{Q}} =1$ in Eq.~\eqref{eq:isd_disp} we obtain the $1/d$ expansion for the energy gap
\begin{equation}
\label{eq:isd_gap}
\frac{\Delta^2}{4 h^2} = 1 - 2 \qq + \frac{\qq^2}{4 d} (4 + 2 \qq) \,.
\end{equation}
As for the dispersion, a $1/d$ expansion for $\Delta$ can be written away from the QPT. In particular, in the small-$\qq$ limit the $1/d$ expansion of $\Delta$ matches with the LCE result \eqref{gap_smallx_tf}. Since the single-particle excitation gap vanishes at the quantum critical point, we can use this criterion to obtain the $1/d$ expansion for the phase boundary to the ferromagnetic phase,
\begin{equation}
\label{eq:isd_pb}
\qqc = \frac{1}{2} + \frac{5}{32 d} +\mathcal{O}\left(\frac{1}{d^2}\right)\,.
\end{equation}

Next, we consider the ground-state energy. In this case, the harmonic approximation itself leads to an expression to order $1/d$, while the diagrammatic contribution starts at $\mathcal{O}(1/d^{2})$. After collecting all the contributions we obtain the following expression for the ground-state energy per site:
\begin{equation}
\label{eq:isd_ge}
\frac{E_{g}}{2 hN} = -\frac{1}{2} - \frac{\qq^2}{8 d} - \frac{7 \qq^4}{64 d^2} +\mathcal{O}\left(\frac{1}{d^3}\right)\,.
\end{equation}
This expression matches with the LCE result Eq.~\eqref{e0_smallx_tf}.

%%%%%%%%%%%%%%%%%%%%%%%%%%%%%%%%%%%%%%%%%%
\subsection{Ferromagnetic phase}
\label{sec:fm}
%%%%%%%%%%%%%%%%%%%%%%%%%%%%%%%%%%%%%%%%%%

We now discuss the symmetry-broken ferromagnetic phase, realized for $\qq > \qqc$. Beyond the quantum critical point, the auxiliary boson introduced earlier is condensed. As a consequence, a suitable reference state in the ordered phase is\cite{sommer}
\begin{equation}
\label{eq:iso_ord}
\Psi_0 = \prod_{i} |\tilde{0} \rangle_i \,, ~~~%\\
%\label{eq:isd_ord1}
| \tilde{0} \rangle_{i} = \frac{| 0 \rangle_i + \lm | T \rangle_i}{\sqrt{1 + \lm^2}} \,.%\,, \\
%| \bar{T} \rangle &= \frac{-\lm | 0 \rangle_i + | T \rangle_i}{\sqrt{1 + \lm^2}} \,.
\end{equation}
%In the spirit of large-$d$, this delivers exact expectation values of local observables in the limit $d \to \infty$.
In this case, $\lm$ is the condensation parameter which takes values between $0$ and $1$ (alternatively, $-1$ for the other $\mathbb{Z}_2$ symmetry related choice of reference state) as a function of the tuning parameter $\qq$. Obviously, $\Psi_0$ in the limit $\lm \to 1$ is the fully polarized ferromagnet. Apart from our reference state we have one more state in the physical Hilbert space, which is orthonormal to the above state and is given by
\begin{equation}
\label{eq:iso_ord2}
| \Tb \rangle_{i} = \frac{-\lm | 0 \rangle_i + | T \rangle_i}{\sqrt{1 + \lm^2}} \,.
\end{equation}
This is a local spin-flip excitation on top of the reference state. Again we introduce auxiliary bosons $\Tb$ such that
\begin{equation}
\label{eq:iso_Tdef}
\Tb_{i}^{\dagger} | \bar{0} \rangle_i = | \Tb \rangle_i \,;~~ \Tb_{i} | \bar{0} \rangle_{i} = 0 \,,
\end{equation}
which obey the usual bosonic commutation relations.
%\begin{equation}
%\label{eq:iso_Tcomm}
%\big[ \Tb_{i}, \Tb_{j}^{\dagger} \big] = \delta_{i,j} \,;~
%\big[ \Tb_{i}, \Tb_{j} \big] = \big[ \Tb_{i}^{\dagger}, \Tb_{j}^{\dagger} \big] = 0 \,.
%\end{equation}
As above, the hard-core constraint is implemented by introducing the projection operator
\begin{equation}
\label{eq:iso_proj}
\Pb_{i} = 1 - \Tb_{i}^{\dagger} \Tb_{i}^{\phantom{\dagger}} \,.
\end{equation}
The spin operators, expressed using the rotated excitation operators $\Tb$, read:
\begin{align}
\label{eq:iso_sx}
\sigma_{i}^{x} &= \frac{1}{1+ \lm^2} \left( (1 - \lm^2)(1 - 2 \Tb_{i}^{\dagger} \Tb_{i}^{\phantom{\dagger}}) - 2\lm (\Tb_{i}^{\dagger} \Pb_{i}^{\phantom{\dagger}} + \Pb_{i} \Tb_{i}) \right) \,, \\
\label{eq:iso_sy}
\sigma_{i}^{y} &= \ii \left( \Pb_{i} \Tb_{i} - \Tb_{i}^{\dagger} \Pb_{i}^{\phantom{\dagger}} \right) \,, \\
\label{eq:iso_sz}
\sigma_{i}^{z} &= \frac{1}{1+ \lm^2} \left( 2\lm(1 - 2 \Tb_{i}^{\dagger} \Tb_{i}^{\phantom{\dagger}}) + (1 - \lm^2) (\Tb_{i}^{\dagger} \Pb_{i}^{\phantom{\dagger}} + \Pb_{i} \Tb_{i}) \right) \,.
\end{align}
Also for $\lm=0$ we recover the spin representation \eqref{eq:sxdef_qp}-\eqref{eq:szdef_qp} in the paramagnetic phase. Parenthetically we note that local spin flips correspond to elementary excitations of the ordered phase only for $d>1$ whereas domain-wall excitations are relevant in $d=1$. Since we are interested in the large-$d$ limit, this is not an issue.

The calculation in the ordered phase is similar to that in the paramagnetic phase, with one important complication: The condensation parameter $\lm$ itself has a $1/d$ expansion, which is another source for $1/d$ corrections to other observables. A detailed discussion regarding the corrections to $\lm$ and its influence can be found in Ref. \onlinecite{Joshi2015b}. We will now straightaway proceed to $1/d$ expansion of observables in this phase. Technical details can be found in Appendix \ref{app:large_d_fm}.

We begin by quoting the single-particle (i.e. spin-wave) dispersion:
\begin{align}
\label{eq:iso_disp}
\frac{\Wkt^{2}}{4 h^{2}} &= 4 \qq^2 - \gk - \frac{1}{d} \frac{1}{128 \qq^{4} (12 \qq^{2} + \gk)} \big[ 2 \gk^3 \nonumber \\
&+ 1536 \qq^{6} (1+\gk) + 4 \gk \qq^{2} (1 - 4 \gk^2) \nonumber \\
&+ 16 \qq^4 (2 \gk^3 + 2 \gk^2 - 16 \gk - 15)\big] \,.
\end{align}
We obtain the excitation-energy gap \mbox{$\tilde{\Delta} = \tilde{\Omega}_{\vec{Q}}$} by substituting $\gamma_{\vec{Q}} = 1$ in the above expression:
\begin{align}
\label{eq:iso_gap}
\frac{\tilde{\Delta}^{2}}{4 h^{2}} &= 4 \qq^2 - 1 - \frac{1}{d} \frac{1}{128 \qq^{4} (12 \qq^{2} + 1)} \big[ 2  \nonumber \\
&- 12 \qq^{2}  - 432 \qq^{4}  %\nonumber \\
+ 3072 \qq^6 \big] +\mathcal{O}\left(\frac{1}{d^2}\right)\,.
\end{align}
Again, demanding that the gap vanishes at the quantum critical point we obtain the same phase boundary \eqref{eq:isd_pb} as before.

Unlike the paramagnetic phase, the ground-state energy in this phase is evaluated only to order $1/d$, because the next order would require corrections to $\lm$ to order $1/d^2$ which are beyond the scope of this work. The $1/d$ expansion of the ground-state energy per site is then given by
\begin{equation}
\label{eq:iso_ge}
\frac{E_g}{2 h N} = -\frac{4 \qq^2 + 1}{4 \qq}  - \frac{1}{d} \frac{1}{256 \qq^3} +\mathcal{O}\left(\frac{1}{d^2}\right)\,.
\end{equation}
Note that at the quantum critical point given by Eq.~\eqref{eq:isd_pb}, above expression matches the ground-state energy \eqref{eq:isd_ge} calculated in the paramagnetic phase to order $1/d$.

The ferromagnetic phase is characterized by the non-zero value of the order parameter, which is magnetization.
%\begin{equation}
%\label{eq:iso_mag_def}
%M = \frac{1}{N} \sum_{i} \langle \sigma_{i}^{z}\rangle \,.
%\end{equation}
It is zero at the quantum critical point, while in the limit $h \to 0$ it takes the value $1$ (or $-1$ for the other choice of symmetry-broken state) corresponding to the fully polarized state. Since the mean-field critical exponent $\beta=1/2$, we have an analytic $1/d$ expansion for the square of magnetization per site as follows:
\begin{equation}
\label{eq:iso_mag}
M^2 = \frac{4 \qq^2 - 1}{4 \qq^2} - \frac{1}{d} \frac{5}{128 \qq^4} +\mathcal{O}\left(\frac{1}{d^2}\right)\,.
\end{equation}
Note that using the condition of vanishing magnetization at the quantum critical point we again get the phase boundary \eqref{eq:isd_pb}.

%%%%%%%%%%%%%%%%%%%%%%%%%%%%%%%%%%%%%%%%%%%%%%%%%%%%%%%%%%%%%%%%%%%%%%%%%%%%%%%%%%%%%%%
%
%%%%%%%%%%%%%%%%%%%%%%%%%%%%%%%%%%%%%%%%%%%%%%%%%%%%%%%%%%%%%%%%%%%%%%%%%%%%%%%%%%%%%%%
\section{Results for the quantum phase transition in arbitrary $d$}
\label{sec:results}
%%%%%%%%%%%%%%%%%%%%%%%%%%%%%%%%%%%%%%%%%%%%%%%%%%%%%%%%%%%%%%%%%%%%%%%%%%%%%%%%%%%%%%%
In the following we use the high-order LCE to estimate the location of the QPT -- as the point where the gap of the symmetric (paramagnetic or field-polarized) phase closes -- for both models and arbitrary $d$. Furthermore, we aim at extracting the correlation-length exponent as well as multiplicative logarithmic corrections at the upper critical dimension $d=3$. Finally, we use the extrapolation of the series expansions about the $d=\infty$ limit to predict the leading coefficients of the full $1/d$ expansion.
%%%%%%%%%%%%%%%%%%%%%%%%%%%%%%%%%%%%%%%%%%%%%%%%%%%%%%%%%%%%%%%%%%%%%%%%%%%%%%%%%%%%%%%
\subsection{Coupled-dimer Heisenberg model}
\label{ssec:results_cd}
\begin{figure}[t]
\begin{center}
\includegraphics[width=8cm]{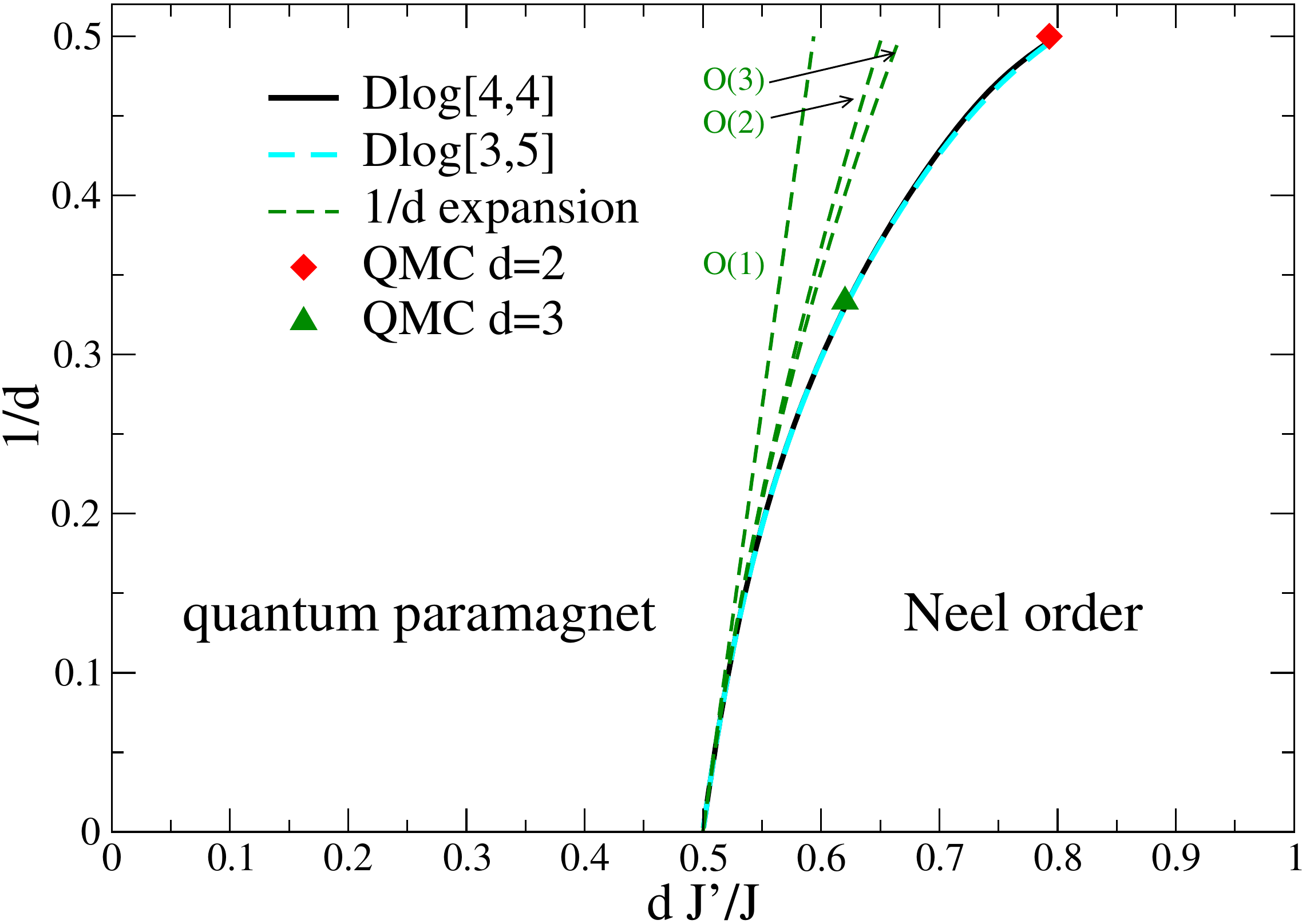}
\caption{Critical point versus inverse dimension $1/d$ for the CDHM on the hypercubic lattice. Red diamond (green triangle) corresponds to the critical value obtained from quantum Monte Carlo simulations in Ref.~\onlinecite{sandvik06} (Ref.~\onlinecite{Qin2015}) for $d=2$ ($d=3$). Green dashed lines represent the estimated full $1/d$ expansion Eq.~\eqref{eq:predicted_dinf_cdhm} up to order $\mathcal{O}(n)$ with $n\in\{1,2,3\}$.}
\label{fig:cpcd}
\end{center}
\end{figure}

%%%%%%%%%%%%%%%%%%%%%%%%%%%%%%%%%%%%%%%%%%%%%%%%%%%%%%%%%%%%%%%%%%%%%%%%%%%%%%%%%%%%%%%
The most reliable way to locate the QPT is the use of Dlog-Pad\'{e} extrapolation Eqs.~\eqref{eq:dlogP1} and \eqref{eq:dlogP2} for the one-triplon gap $\Delta^{\rm CDHM}$ given in Eq.~\eqref{gap_smallx_cd}. To this end we set the inverse of dimension $1/d$ to fixed values in $ [0,0.5]$ and extract the critical point $\x_{\rm c}$. Representative Dlog-Pad\'{e} extrapolants as well as known literature values for $d=2$ and $d=3$ are displayed in Fig.~\ref{fig:cpcd}.
%
%%%%%%%%%%%%%%%%%%%%%%%%%%%%%%%%%%%%%%%%%%%%%%%%%%%%%%%%%%%%%%%%%%%%%%%%%%%%%%%%%%%%%%%
%%%%%%%%%%%%%%%%%%%%%%%%%%%%%%%%%%%%%%%%%%%%%%%%%%%%%%%%%%%%%%%%%%%%%%%%%%%%%%%%%%%%%%%

One observes that the extrapolants vary smoothly with $1/d$ and agree well with the literature values for $d=2$ and $d=3$ as well as with the known $d=\infty$ limit. As one important example, where especially no series expansion for fixed dimensions is available, let us focus on $d=3$.
 If we average over all Dlog-Pad\'{e} extrapolants $P[L/M]_{D}$ with $L+M=8$ and $L,M\leq 2$, we obtain $\qqc^{\rm CDHM} =0.6206(5)$ which is in excellent agreement with $\qqc^{\rm CDHM} = 0.620$ from quantum Monte Carlo simulations \cite{Qin2015}.
%
%%%%%%%%%%%%%%%%%%%%%%%%%%%%%%%%%%%%%%%%%%%%%%%%%%%%%%%%%%%%%%%%%%%%%%%%%%%%%%%%%%%%%%%
\begin{figure}[ht]
\begin{center}
\includegraphics[width=8cm]{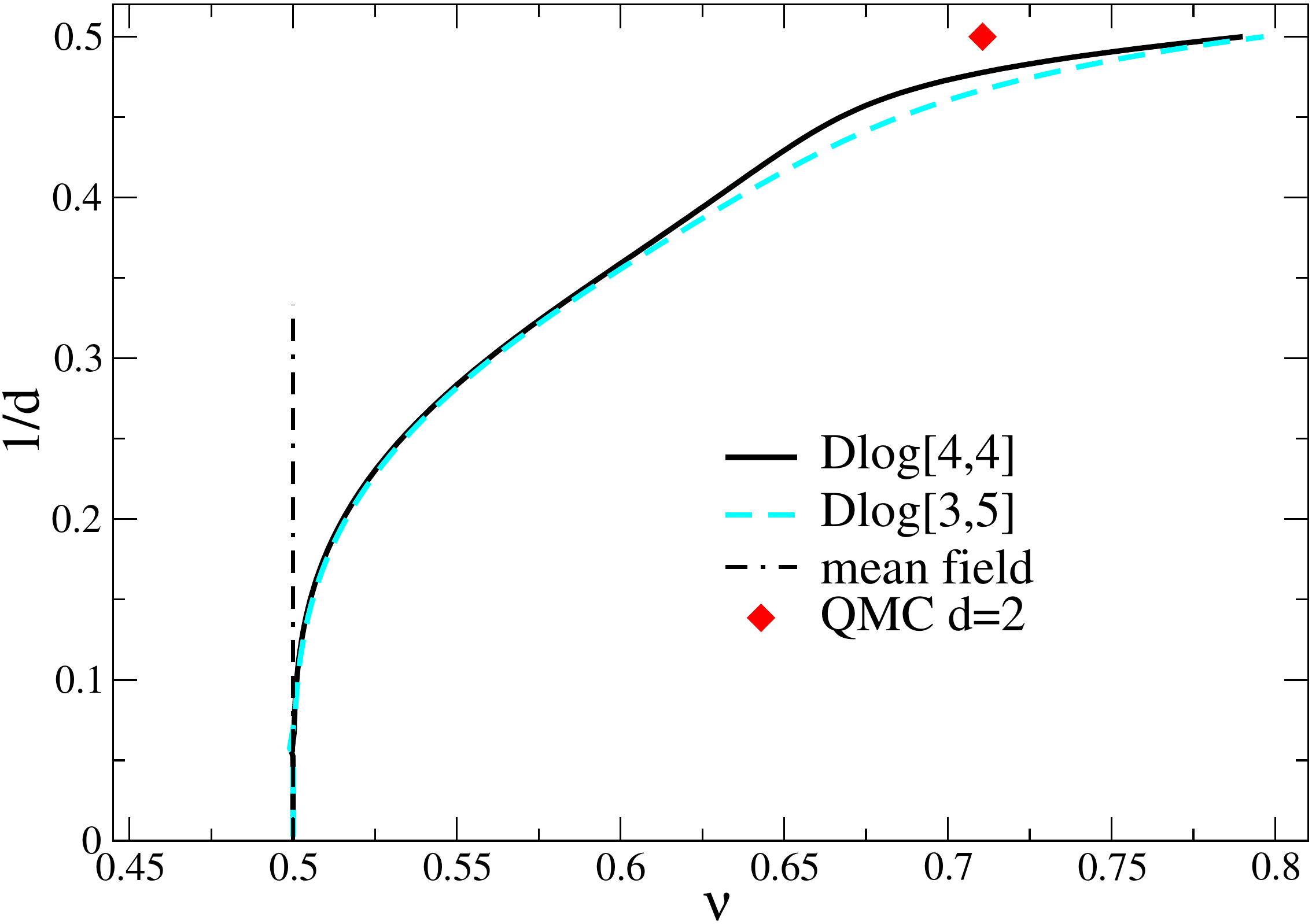}
\caption{Critical exponent $\nu$ versus inverse dimension $1/d$ for the CDHM on the hypercubic lattice. Dot-dashed line indicates the mean-field exponent $\nu=1/2$ and the red diamond corresponds to the critical exponent obtained from quantum Monte Carlo simulations in Ref.~\onlinecite{campostrini2002} for $d=2$.}
\label{fig:cecd}
\end{center}
\end{figure}
%%%%%%%%%%%%%%%%%%%%%%%%%%%%%%%%%%%%%%%%%%%%%%%%%%%%%%%%%%%%%%%%%%%%%%%%%%%%%%%%%%%%%%%

Next we investigate the behavior of the critical exponent $\nu$ of the one-triplon gap (recall $z=1$) close to the quantum critical line as a function of $1/d$ using Eq.~\eqref{extract_exponent}. The corresponding results are given in Fig.~\ref{fig:cecd}.

Above the upper critical dimension, $d\geq 3$, mean-field behavior with $\nu=1/2$ is expected, whereas for $3>d\geq 2$ one expects a continuously varying critical exponent. In particular, $\nu=0.7113(10)$ for $d=2$ \cite{campostrini2002}. Semi-quantitatively, this behavior is reproduced by the Dlog-Pad\'{e} extrapolants. However, for $d=2$ the extrapolated critical exponent overshoots the correct value $\nu\approx 0.71$. This difference can be attributed to the maximal order nine of our series expansion. Indeed, if one performs high-order series expansions for fixed dimension $d=2$, one is able to reach order 11 which yields a critical exponent much closer to $\nu\approx 0.71$ \cite{Weihong97}.

The other issue is the behavior close to the upper critical dimension $d=3$. By construction, the series expansion cannot be expected to give a constant value for $d\geq3$; it will rather yield a smooth behavior $\nu(d)$. Nevertheless, it may be surprising that sizeable deviations from mean-field behavior are visible already for $d\lesssim 6$. We attribute this behavior to a combination of small scaling dimension and large prefactor of the leading irrelevant perturbation for $d\gtrsim 3$ which in turn spoils the estimate of the critical exponent from series expansions.

\begin{figure}[bt]
\begin{center}
\includegraphics[width=8cm]{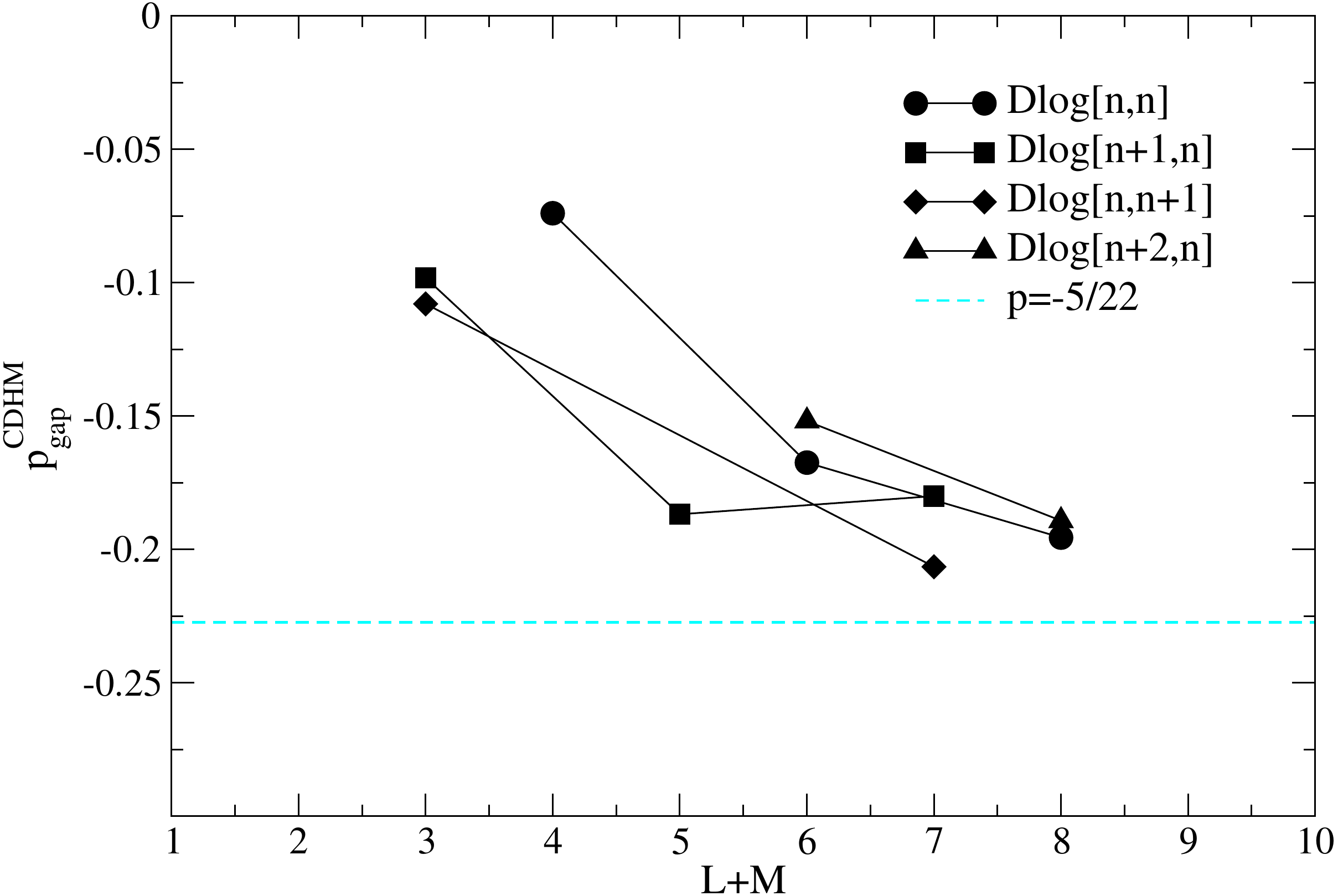}
\caption{Multiplicative logarithmic correction $p^{\rm CDHM}_{\rm gap}$ versus perturbative order $L+M$ used in the Dlog-Pade\'{e} extrapolation Eq.~\eqref{biasp} with $\qqc^{\rm CDHM} =0.6206$ for the CDHM on the hypercubic lattice. Extrapolants with constant $c=L-M$ with $c\in\{-1,0,1,2\}$ are shown with the same black symbols. Dashed line indicates the exact exponent\cite{kenna2012} $(-5/22)$.}
\label{fig:logcd}
\end{center}
\end{figure}

As described in subsection \ref{ssec:extrapol}, it is also possible to estimate the multiplicative logarithmic corrections at $d=3$ using Eq.~\eqref{biasp}. To this end we fix the critical exponent to $1/2$ and the critical point to $\qqc^{\rm CDHM} =0.6206$. The corresponding values of $p^{\rm CDHM}_{\rm gap}$ as a function of perturbative order $L+M$ are displayed in Fig.~\ref{fig:logcd} together with the exact value\cite{kenna2012,normand2016} $(-5/22)$ derived from perturbative renormalization-group (RG) calculations. Averaging over Dlog-Pad\'{e} extrapolants with $L+M\geq 7$ and $M-L\leq 2$ yields $p^{\rm CDHM}_{\rm gap}=-0.19(2)$. Let us stress that these value are quite sensitive to the critical value $\qqc^{\rm CDHM}$ entering the extrapolation. If one uses the value $\qqc^{\rm CDHM} =0.620$ from quantum Monte Carlo simulations \cite{Qin2015}, the value for the multiplicative logarithmic correction changes to \mbox{$p^{\rm CDHM}_{\rm gap}=-0.18(3)$}. Nevertheless, as can be seen from Fig.~\ref{fig:logcd}, the extrapolation tend to decrease with increasing perturbative order so that our results are in full agreement with the exact value $(-5/22)$.

%
%%%%%%%%%%%%%%%%%%%%%%%%%%%%%%%%%%%%%%%%%%%%%%%%%%%%%%%%%%%%%%%%%%%%%%%%%%%%%%%%%%%%%%%
%%%%%%%%%%%%%%%%%%%%%%%%%%%%%%%%%%%%%%%%%%%%%%%%%%%%%%%%%%%%%%%%%%%%%%%%%%%%%%%%%%%%%%%

\subsection{Transverse-field Ising model}
\label{ssec:results_tfim}
\begin{figure}[bt]
\begin{center}
\includegraphics[width=8cm]{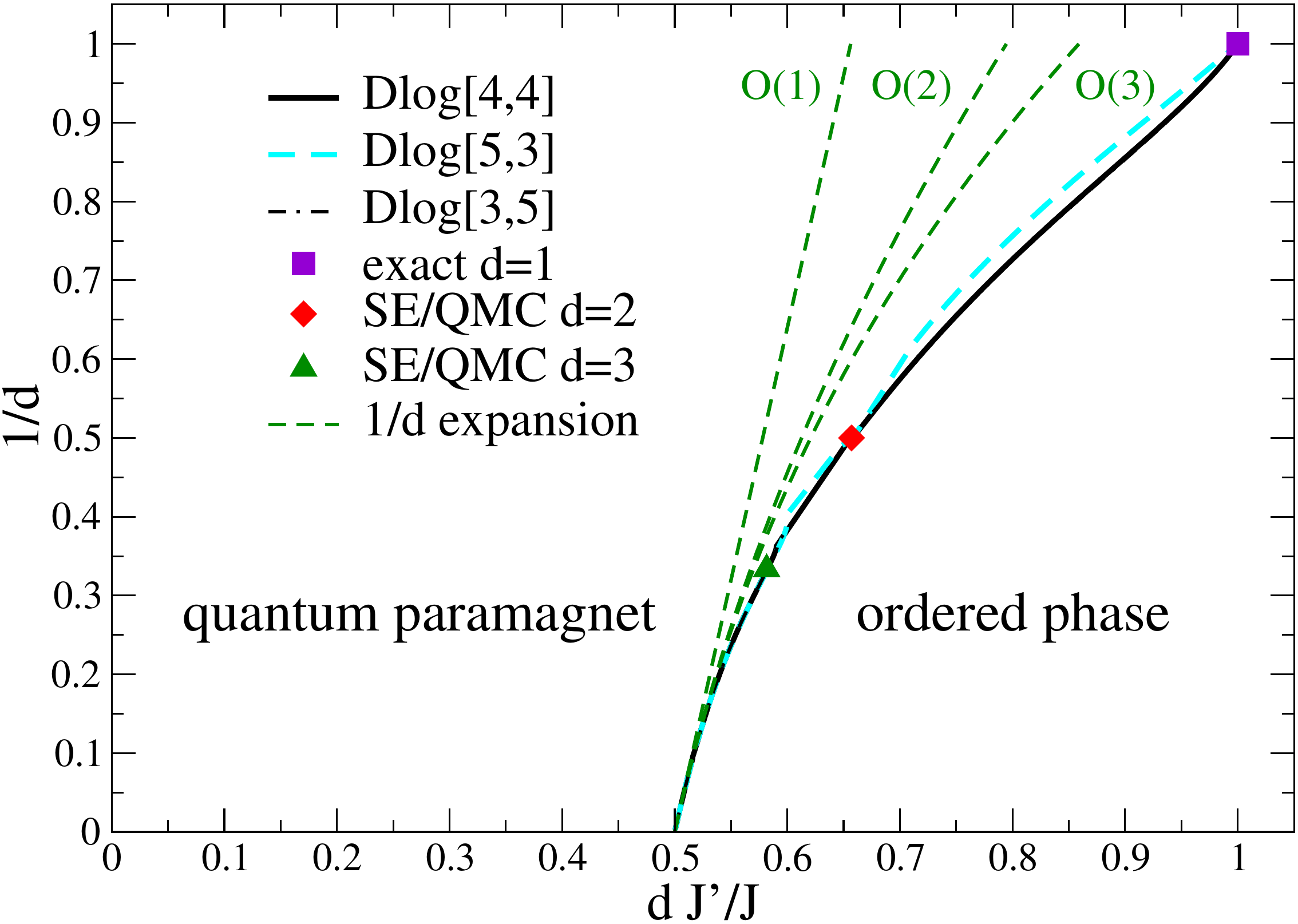}
\caption{Critical point versus inverse dimension $1/d$ for the TFIM on the hypercubic lattice. Red diamond (green triangle) corresponds to the critical value obtained from series expansion (SE) in Ref.~\onlinecite{He1990} (Ref.~\onlinecite{Zheng1994}) for $d=2$ ($d=3$) and quantum Monte Carlo (QMC) simulations from Ref.~\onlinecite{deng_qmc3d} (the two techniques fully agree on the displayed scale and are therefore shown together as a single symbol). Blue square corresponds to the exact solution for $d=1$ \cite{Pfeuty1970}. Green dashed lines represent the estimated full $1/d$ expansion Eq.~\eqref{eq:predicted_dinf_tfim} up to order $\mathcal{O}(n)$ with $n\in\{1,2,3\}$.
}
\label{fig:cptfim}
\end{center}
\end{figure}

Let us turn to the TFIM on the hypercubic lattice. For this model one expects a second-order QPT for all dimensions $d\geq 1$, separating the gapped polarized phase present at large fields from the gapped symmetry-broken phase for dominating Ising exchange. We have again used DlogPad\'{e} extrapolation to estimate the location of the QPT as a function of $1/d$. DlogPad\'{e} extrapolants as well as the known results from literature are shown in Fig.~\ref{fig:cptfim}.
%
%%%%%%%%%%%%%%%%%%%%%%%%%%%%%%%%%%%%%%%%%%%%%%%%%%%%%%%%%%%%%%%%%%%%%%%%%%%%%%%%%%%%%%%
%%%%%%%%%%%%%%%%%%%%%%%%%%%%%%%%%%%%%%%%%%%%%%%%%%%%%%%%%%%%%%%%%%%%%%%%%%%%%%%%%%%%%%%

\begin{figure}[t]
\begin{center}
\includegraphics[width=8cm]{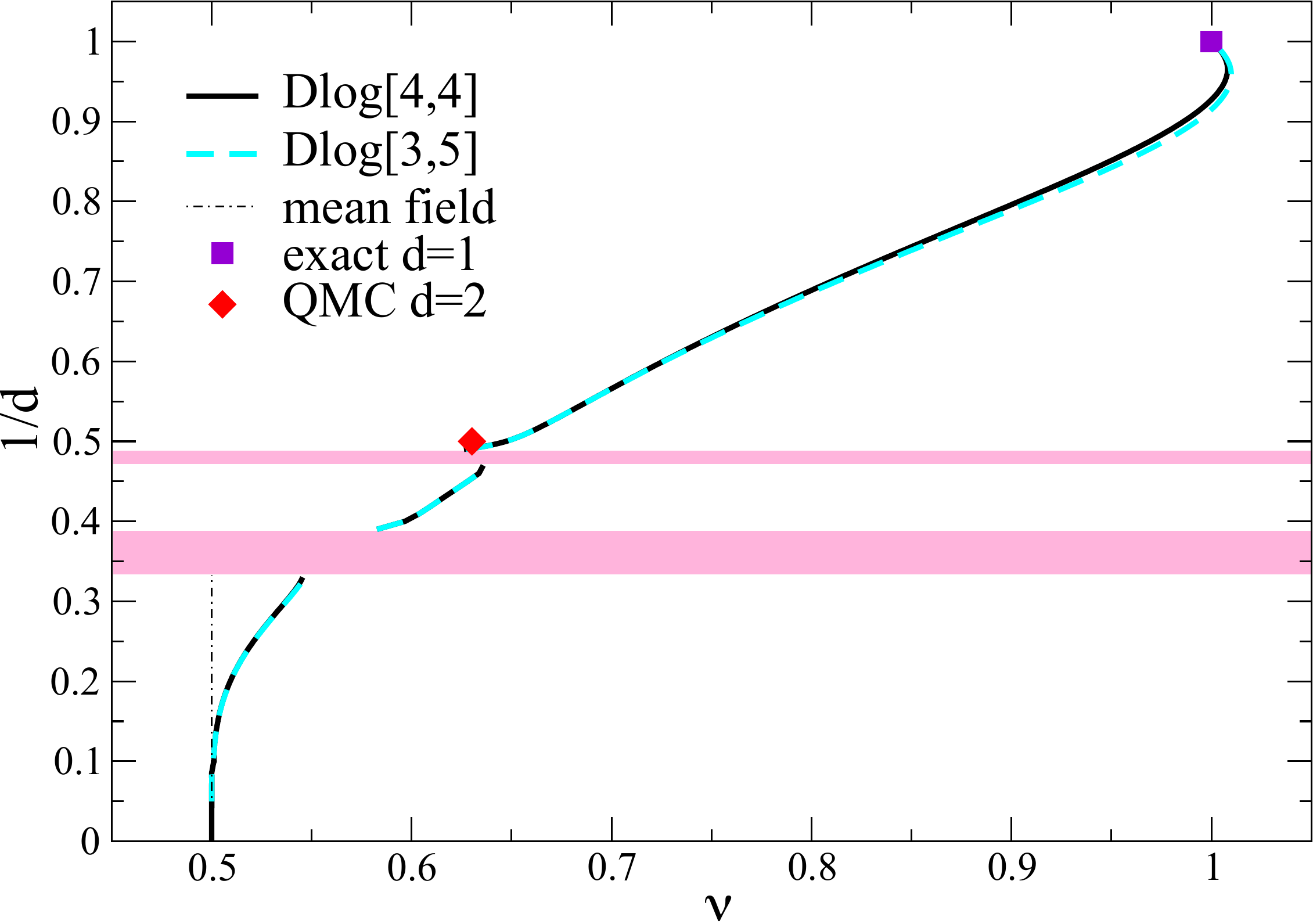}
\caption{Critical exponent $\nu$ versus dimension for the TFIM on the hypercubic lattice. Dot-dashed line indicates the mean-field exponent $\nu=1/2$. The red diamond corresponds to the critical exponent obtained from quantum Monte Carlo simulations in Ref.~\onlinecite{Bloete1995} for $d=2$. The blue square illustrates the exact value for $d=1$ \cite{Pfeuty1970}. In the shaded regions no consistent extrapolation has been achieved due to spurious poles in the Dlog-Pad\'{e} extrapolation.}
\label{fig:cetfim}
\end{center}
\end{figure}

Overall, the series expansion yields convincing results capturing quantitatively the exactly
 known cases $1/d=1$ \cite{Pfeuty1970} and $1/d=0$ as well as the estimates of series expansions with higher maximal order for fixed dimensions $d=2$ and $d=3$. The associated critical exponent $\nu$ ($z=1$ for the TFIM) as a function of $1/d$ is displayed in Fig.~\ref{fig:cetfim}. In certain regimes in $1/d$ (the two shaded areas in Fig.~\ref{fig:cetfim}) we found no consistent Dlog-Pad\'{e} extrapolation, since there are two poles close to each other on the real axis in the denominator of Eq.~\eqref{dx} which spoil the extrapolation scheme. Similarly to the CDHM, one observes deviations from mean-field behavior for $3<d\lesssim 6$. As above, we believe that these are caused by large subleading corrections with small scaling dimension.

For the TFIM we do not analyze the multiplicative logarithmic corrections the the gap behavior, since Ref.~\onlinecite{Zheng1994} already performed such an analysis with the high-order series expansion of order 13 for $d=3$. This gives \mbox{$p^{\rm TFIM}_{\rm gap}=-0.143(5)$} which is consistent with the exact value\cite{Larkin1969,Brezin1973,Wegner1973,Zheng1994} $(-1/6)$ from perturbative RG.
%
%%%%%%%%%%%%%%%%%%%%%%%%%%%%%%%%%%%%%%%%%%%%%%%%%%%%%%%%%%%%%%%%%%%%%%%%%%%%%%%%%%%%%%%
%%%%%%%%%%%%%%%%%%%%%%%%%%%%%%%%%%%%%%%%%%%%%%%%%%%%%%%%%%%%%%%%%%%%%%%%%%%%%%%%%%%%%%%

%%%%%%%%%%%%%%%%%%%%%%%%%%%%%%%%%%%%%%%%%%%%%%%%%%%%%%%%%%%%%%%%%%%%%%%%%%%%%%%%%%%%%%%
\subsection{Large-$d$ limit}
%%%%%%%%%%%%%%%%%%%%%%%%%%%%%%%%%%%%%%%%%%%%%%%%%%%%%%%%%%%%%%%%%%%%%%%%%%%%%%%%%%%%%%%
Up to now, we have studied the one-particle gap for both models by fixing the dimension $d$ and  using DlogPad\'{e} extrapolation to extract critical points and associated critical exponents. In this part we use the extrapolation of the high-order series expansion in $\x$ to get an estimate for the quantum critical line about the $1/d=0$ limit in powers of $1/d$, i.e.~we are interested in determining the coefficients $c_n$ in
\begin{equation}
 \label{eq:poly}
 \qqc=c_0+c_1\,\left(\frac{1}{d}\right)+ c_2\,\left( \frac{1}{d} \right)^2+ c_3\,\left( \frac{1}{d}\right)^3 \ldots \quad .
\end{equation}
In a first approach we used various DlogPad\'{e} extrapolants for small values of the inverse dimension $1/d=0,\epsilon,2\epsilon,\ldots$ and we fitted the polynomial \eqref{eq:poly} to determine the coefficients $c_n$. We observed that the coefficients $c_n$ converged reliably under the variation of $\epsilon$ and for different DlogPad\'{e} extrapolants, and our numerical estimates for $c_0$ and $c_1$ approach the exact value from the analytic $1/d$ expansion.

The latter findings motivate the second approach, where the $c_n$ are obtained as fractions from the DlogPad\'{e} extrapolation. To this end we did not fix the value of the dimension in the DlogPad\'{e} extrapolation, but we keep it general when calculating the Pad\'{e} of the logarithmic derivative Eq.~\eqref{eq:dlogP2}. In a next step we determine the pole of the denominator,  corresponding to the critical point, as a Taylor series in $1/d$ which exactly yields a polynomial of the form Eq.~\eqref{eq:poly}. And indeed, we find that the leading coefficients $c_0$ and $c_1$ do not depend on the specific DlogPad\'{e} extrapolant used and, more importantly, the values correspond exactly to the values from the analytical $1/d$ expansion. In contrast, the higher orders $c_2$ and $c_3$ do depend on the specific extrapolant, but the obtained values are very close to each other numerically and consistent with the values from the first approach.

One then obtains the following expressions for the $1/d$ expansion up to order three
\begin{equation}
 \label{eq:predicted_dinf_cdhm}
 \qqc^{\rm CDHM}=\frac{1}{2}+\frac{3}{16}\,\left(\frac{1}{d}\right)+ 0.2311\,\left( \frac{1}{d} \right)^2+ 0.1233\,\left( \frac{1}{d}\right)^3 \ldots
\end{equation}
for the CDHM and
\begin{equation}
\label{eq:predicted_dinf_tfim}
 \qqc^{\rm TFIM}=\frac{1}{2}+\frac{5}{32}\,\left(\frac{1}{d}\right)+ 0.1383\,\left( \frac{1}{d} \right)^2+ 0.0643\,\left( \frac{1}{d}\right)^3 \ldots
\end{equation}
for the TFIM. These results are also illustrated in the phase diagram of the CDHM (TFIM) in Fig.~\ref{fig:cpcd} (Fig.~\ref{fig:cptfim}). One observes that the full $1/d$ expansion for both models is well behaved in the sense that all deduced coefficients are positive and one therefore approaches the correct quantum critical line at finite dimensions steadily with increasing order in $1/d$.
%
%%%%%%%%%%%%%%%%%%%%%%%%%%%%%%%%%%%%%%%%%%%%%%%%%%%%%%%%%%%%%%%%%%%%%%%%%%%%%%%%%%%%%%%
\section{Conclusion}
%%%%%%%%%%%%%%%%%%%%%%%%%%%%%%%%%%%%%%%%%%%%%%%%%%%%%%%%%%%%%%%%%%%%%%%%%%%%%%%%%%%%%%%
\label{sec:conclusion}

We used high-order series expansions for arbitrary dimensions $d$ to study the dimensional dependence of the quantum critical line for two paradigmatic models in quantum magnetism, namely the transverse-field Ising model and the coupled-dimer Heisenberg model on the hypercubic lattice. In both cases we reached order 9 perturbation theory about the decoupled-dimer (high-field) limit for the ground-state energy per dimer (site) as well as for the one-particle gap for general dimension $d$. This is achieved by a full graph decomposition in linked graphs keeping in mind that the dimension $d$ enters only in the combinatorical embedding factor of the graphs.

We focused on analyzing for both models the behavior of the one-particle gap about the decoupled-dimer (high-field) limit. In both cases the continuous critical line of the LCE is in quantitative agreement with the known results for fixed finite dimensions as well as with the leading orders of the $1/d$ expansion. For the CDHM, the series expansion for fixed dimension $d=3$ were to the best of our knowledge unknown and the extrapolated critical point is found to be in quantitative agreement with QMC simulations \cite{Qin2015}. Furthermore, we also extracted the multiplicative logarithmic correction for this case which is fully consistent with the value from perturbative renormalization group calculations.

Finally, we used the LCEs to predict the coefficients of the full $1/d$ expansion of the quantum critical line. It is found that the series in $1/d$ up to order 3 has only positive coefficients for both models and therefore approach monotonously the correct values. These subleading corrections are however not small. In our opinion the most remarkable finding is that the Dlog-Pad\'{e} extrapolation of the gap series yields exactly the leading coefficients $(1/d)^0$ and $(1/d)^1$ of the full $1/d$ expansion. It would be interesting to study this issue in other models in future studies.

%%%%%%%%%%%%%%%%%%%%%%%%%%%%%%%%%%%%%%%%%%%%%%%%%%%%%%%%%%%%%%%%%%%%%%%%%

\acknowledgments

We thank B. Normand for helpful discussions concerning logarithmic corrections. This work was in part supported by the Helmholtz Virtual Institute ``New states of matter and their excitations'' as well as from the Deutsche Forschungsgemeinschaft (DFG) via grants SCHM 2511/9-1 and SFB 1143.

%%%%%%%%%%%%%%%%%%%%%%%%%%%%%%%%%%%%%%%%%%%%%%%%%%%%%%%%%%%%%%%%%%%%%%%%%%%%%%%

%%%%%%%%%%%%%%%%%%%%%%%%%%%%%%%%%%%%%%%%%%%%%%%%%%%%%%%%%%%%%%%%%%%%%%%%%%%%%%%%%%%%%%%
\appendix

%%%%%%%%%%%%%%%%%%%%%%%%%%%%%%%%%%%%%%%%%%%%%%%%%%%%%%%%%%%%%%%%%%%%%%%%%%%%%%%
\section{$1/d$ expansion details: Quantum paramagnetic phase}
\label{app:large_d_qp}
%%%%%%%%%%%%%%%%%%%%%%%%%%%%%%%%%%%%%%%%%%%%%%%%%%%%%%%%%%%%%%%%%%%%%%%%%%%%%%%

Inserting Eqs.~\eqref{eq:sxdef_qp} - \eqref{eq:szdef_qp} in the spin Hamiltonian \eqref{tfim} we obtain the following interacting boson Hamiltonian
\begin{align}
\label{eq:realham_qp}
\mathcal{H} = &-J \sum_{\langle ij \rangle} \left( P_{i}T_{i}P_{j}T_{j} + T_{i}^{\dagger} P_{i}^{\phantom{\dagger}} P_{j}^{\phantom{\dagger}} T_{j}^{\phantom{\dagger}} + {\rm h.c.} \right) \nonumber \\
&- h \sum_{i} \left( 1 - 2 T_{i}^{\dagger} T_{i}^{\phantom{\dagger}}  \right) \,.
\end{align}
Further, inserting the explicit expression for the projection operator \eqref{eq:proj_qp} we can write down the various pieces $\mathcal{H}_{n}$, where $n$ denotes the number of $T$ operators. In this case, we have only even-ordered $\mathcal{H}_{n}$. It turns out that to order $1/d$ calculation we need terms only upto $\mathcal{H}_{4}$. We therefore present the explicit expressions of the relevant terms below:
\begin{align}
\label{eq:h0_qp}
\mathcal{H}_0 &= - N h \,, ~~~ \text{$N$ is the number of lattice sites} \,,\\
\label{eq:h2_qp}
\mathcal{H}_2 &= - J \sum_{\langle ij \rangle} \left(  T_{i} T_{j} + T_{i}^{\dagger} T_{j}^{\phantom{\dagger}}  + {\rm h.c.} \right)
+ 2 h \sum_{i} T_{i}^{\dagger} T_{i}^{\phantom{\dagger}}  \,, \\
\label{eq:h4_qp}
\mathcal{H}_4 &= 2 J \sum_{\langle ij \rangle} \bigg[ T_{i}^{\dagger} T_{i}^{\phantom{\dagger}}  T_{i}^{\phantom{\dagger}}  T_{j}^{\phantom{\dagger}}  + T_{i}^{\dagger} T_{i}^{\dagger} T_{i}^{\phantom{\dagger}}  T_{j}^{\phantom{\dagger}}  + {\rm h.c.} \bigg] \,.
\end{align}
%%%%%%%%%%%%%%%

Let us first solve the bilinear part \eqref{eq:h2_qp}, which we call the {\em harmonic approximation}. Utilizing the lattice translation symmetry, it is convenient to work in the Fourier space by introducing
\begin{equation}
\label{eq:Tk}
T_{i} = \frac{1}{\sqrt{N}} \sum_{\kk} T_{\kk} e^{-\ii \kk \cdot \vec{r}_{i}} \,.
\end{equation}
%such that
%\begin{equation}
%\label{eq:Tkcomm}
%\big[ T_{\kk}, T_{\vec{k}'}^{\dagger} \big] = \delta_{\kk,\vec{k}'} \,,~~~
%\big[ T_{\kk}, T_{\vec{k}'} \big] = \big[ T_{\kk}^{\dagger}, T_{\vec{k}'}^{\dagger} \big] = 0 \,.
%\end{equation}
%%%%%%%
In terms of the Fourier transformed operator, $T_{\kk}$, the bilinear part of the Hamiltonian is given by
\begin{equation}
\label{eq:h2k_qp}
\mathcal{H}_2 = \sum_{\kk} \left[ A_{\kk} T_{\kk}^{\dagger} T_{\kk}^{\phantom{\dagger}}  + \frac{B_{\kk}}{2} \left( T_{\kk} T_{-\kk} + {\rm h.c.}\right) \right]
\end{equation}
where
\begin{equation}
\label{eq:bareAB_qp}
A_{\vec k} = 2 h + B_{\vec k}\,,~~
B_{\vec k} = -\gk \qq \,.
%\qq = \frac{Jd}{h}
\end{equation}
The above Hamiltonian piece can be diagonalized by introducing bosonic Bogoliubov transformation,
\begin{equation}
\label{eq:bogol}
T_{\vec{k}} =
u_{\vec{k}}\tau_{\vec{k}}+v_{\vec{k}}\tau^\dagger_{-\vec{k}} \,.
\end{equation}
Here $u_{\kk}$ and $v_{\kk}$ are the Bogoliubov coefficients such that $u_{\kk}^{2} - v_{\kk}^{2} =1$, and the $\tau$ operators obey the usual bosonic commutation relations. Within the harmonic approximation, the mode energy is given by
\begin{equation}
\label{eq:om0}
\wk = \sqrt{A_{\kk}^{2} - B_{\kk}^{2}} = 2 h \sqrt{1 - 2 \gk \qq} \,,
\end{equation}
and the Bogoliubov coefficients are given by
\begin{equation}
\label{eq:isd_bogco}
u_{\kk}^{2}, v_{\kk}^{2} = \frac{1}{2} \left( \frac{A_{\kk}}{\wk} \pm 1 \right) \,; ~~~
u_{\kk} v_{\kk} = - \frac{B_{\kk}}{2 \wk} \,.
\end{equation}

Having solved the bilinear piece of the Hamiltonian, we shall treat it as the unperturbed part and take into account the interaction terms perturbatively.
%We will see that the self-energy from the interaction terms is suppressed in powers of $1/d$. Hence the corrections to observables can be arranged in a power series in $1/d$.
To set-up the method, we must first normal-order our Hamiltonian with respect to the $\tau$ operators. Upon normal ordering, $\mathcal{H}_4$ will generate additional bilinear terms which are expressed below:
\begin{equation}
\label{eq:isd_h2pb}
\mathcal{H}'_{2b} = \sum_{\kk} \bigg[ C_{\kk} \tau_{\kk}^{\dagger} \tau_{\kk}^{\phantom{\dagger}}  + \frac{D_{\kk}}{2} (\tau_{\kk} \tau_{-\kk} + {\rm h.c.})\bigg]
\end{equation}
where
\begin{align}
\label{is_dis_c}
C_{\kk} &= 4 \qq h \big[ (u_{\kk}^{2} + v_{\kk}^{2}) (\gk R_{1} + 2 \gk R_{2} + 2R_{3}) \nonumber \\
&+ 2 u_{\kk} v_{\kk} (\gk R_{1} + 2 \gk R_{2} + R_{3}) \big] \,, \\
\label{is_dis_d}
D_{\kk} &= 4 \qq h \big[ (u_{\kk}^{2} + v_{\kk}^{2}) (\gk R_{1} + 2 \gk R_{2} + R_{3}) \nonumber \\
&+ 2 u_{\kk} v_{\kk} (\gk R_{1} + 2 \gk R_{2} + 2R_{3}) \big] \,.
\end{align}
Following are required expressions of R's to order $1/d$ in the large-$d$ limit:
\begin{align}
\label{is_dis_r1}
R_1 &= \frac{1}{N} \sum_{\vec k} u_{\vec k} v_{\vec k} ~~~= \frac{\qq^2}{4 d} + \mathcal{O}(d^{-2}) \,, \\
%%%%%%%
\label{is_dis_r2}
R_2 &= \frac{1}{N} \sum_{\vec k} v^2_{\vec k} ~~~~~~= \frac{\qq^2}{8 d} + \mathcal{O}(d^{-2}) \,, \\
%%%%%%%
\label{is_dis_r3}
R_3 &= \frac{1}{N} \sum_{\vec k} \gk u_{\vec k} v_{\vec k} = \frac{\qq}{4 d} + \mathcal{O}(d^{-2}) \,, \\
%%%%%%%%
\label{is_dis_r4}
R_4 &= \frac{1}{N} \sum_{\vec k} \gamma_{\vec k} v^2_{\vec k} ~~~\,= \mathcal{O}(d^{-2}) \,.
\end{align}
Thus the normal ordered bilinear piece is sum of the unperturbed part and the above contribution: $\mathcal{H}'_{2} = \mathcal{H}'_{2a} + \mathcal{H}'_{2b}$ where,
\begin{equation}
\label{isd_h2pa}
\mathcal{H}'_{2a} = \sum_{\kk} \wk \tau_{\kk}^{\dagger} \tau_{\kk}^{\phantom{\dagger}}
\end{equation}
is the unperturbed piece.

Now we quote the normal ordered quartic term:
\begin{align}
\label{is_dis_h4p}
\mathcal{H}'_4 &= \frac{1}{N} \sum_{1234}
\big[\delta_{1+2+3+4} \vqad
(\tau^\dagger_{1}\tau^\dagger_{2}\tau^\dagger_{3}\tau^\dagger_{4} +
\tau_{1}\tau_{2}\tau_{3}\tau_{4})   \nonumber \\
&~~~~~~~~~~~
+ \delta_{1+2-3-4} (\vqbd
\tau^\dagger_{1}\tau^\dagger_{2}\tau_{3}^{\phantom{\dagger}} \tau_{4}^{\phantom{\dagger}}
+\vqcd \tau^\dagger_{1}\tau^\dagger_{2}\tau_{3}^{\phantom{\dagger}} \tau_{4}^{\phantom{\dagger}} )
\nonumber \\
&~~~~~~~~~~~+\delta_{1+2+3-4} \vqdd
(\tau^\dagger_{1}\tau^\dagger_{2}\tau^\dagger_{3}\tau_{4}^{\phantom{\dagger}}  +
\tau^\dagger_{4} \tau_{3}^{\phantom{\dagger}} \tau_{2}^{\phantom{\dagger}} \tau_{1}^{\phantom{\dagger}} )\big] \,,
\end{align}
with the relevant vertex functions given by
\begin{align}
\vqad &= 2 \qq h \big[ \gamma_{4} u_{1} v_{2} v_{3} v_{4} + \gamma_{4} v_{1} u_{2} u_{3} u_{4} \nonumber \\
&~~~~~~~~~~~+ (\gamma_{1} + \gamma_{4}) u_{1} u_{2} v_{3} v_{4} \big] \,, \\
%%%%%%%%
\vqdd &= 2 \qq h \big[ (2 \gamma_{3} + \gamma_{4}) u_{1} v_{2} v_{3} u_{4} + (2 \gamma_{3} + \gamma_{4}) v_{1} u_{2} u_{3} v_{4} \nonumber \\
&~~~~~~~~~~~+ \gamma_{3} v_{1} v_{2} v_{3} v_{4} %\nonumber \\
+ \gamma_{3} u_{1} u_{2} u_{3} u_{4}  \nonumber \\
&~~~~~~~~~~~+ (2 \gamma_{1} + \gamma_{3} + \gamma_{4}) u_{1} u_{2} v_{3} u_{4} \nonumber \\
&~~~~~~~~~~~+ (\gamma_{1} + 2 \gamma_{3} + \gamma_{4}) u_{1} v_{2} v_{3} v_{4} \big] \,.
\end{align}
The self-energy diagrams constructed using these quartic vertices along with the bilinear vertices coming from \eqref{eq:isd_h2pb} are suppressed as $1/d$. After inserting the self-energy in the Dyson equation and identifying the pole of the Green's functions, we obtain the dispersion \eqref{eq:isd_disp}. The diagrammatic expansion for the ground-state energy works along similar lines, and we refer to Ref. \onlinecite{Joshi2015} for more details.

%%%%%%%%%%%%%%%%%%%%%%%%%%%%%%%%%%%%%%%%%%%%%%%%%%%%%%%%%%%%%%%%%%%%%%%%%%%%%%%
\section{$1/d$ expansion details: Ferromagnetic phase}
\label{app:large_d_fm}
%%%%%%%%%%%%%%%%%%%%%%%%%%%%%%%%%%%%%%%%%%%%%%%%%%%%%%%%%%%%%%%%%%%%%%%%%%%%%%%

Using Eqs.~\eqref{eq:iso_sx} - \eqref{eq:iso_sz}, the Hamiltonian in the ordered phase is
\begin{widetext}
\begin{align}
\label{eq:iso_ham}
\mathcal{H} = &- \frac{J}{ (1+\lm^2)^2} \sum_{\langle ij \rangle} \bigg[ (1-\lm^2)^{2} \left( \Pb_{i} \Tb_{i} \Pb_{j} \Tb_{j} +
\Tb_{i}^{\dagger} \Pb_{i} \Pb_{j} \Tb_{j} + {\rm h.c.} \right)
+ 4 \lm^2 \left( 1 - 2\Tb_{j}^{\dagger} \Tb_{j} - 2\Tb_{i}^{\dagger} \Tb_{i} + 4 \Tb_{i}^{\dagger} \Tb_{i} \Tb_{j}^{\dagger} \Tb_{j}\right) \nonumber \\
&~~~~~~~~~~~~~~~~~~~~~~~~~~~~~+ 2\lm(1-\lm^2) \left( \Tb_{i}^{\dagger} \Pb_{i} -
2 \Tb_{i}^{\dagger} \Pb_{i} \Tb_{j}^{\dagger} \Tb_{j}  +
\Tb_{j}^{\dagger} \Pb_{j} -
2 \Tb_{j}^{\dagger} \Pb_{j} \Tb_{i}^{\dagger} \Tb_{i} + {\rm h.c.} \right)
\bigg] \nonumber \\
&- \frac{h}{1+\lm^2} \sum_{i} \bigg[ (1-\lm^2)(1 - 2\Tb_{i}^{\dagger}\Tb_{i}) - 2\lm (\Tb_{i}^{\dagger} \Pb_{i} + {\rm h.c.} ) \bigg] \,.
\end{align}
\end{widetext}
To order $1/d$ we need only terms up to fourth order in $\Tb$ operators. Inserting the explicit expression of the projector operator \eqref{eq:iso_proj}, we can write the relevant pieces in the Hamiltonian as follows:
\begin{align}
\label{eq:iso_h0}
\mathcal{H}_0 &= -\frac{4 N \qq h \lm^2}{(1+ \lm^2)^2} - \frac{Nh (1- \lm^2)}{1+ \lm^2} \,, \\
%%%%%
\label{eq:iso_h1}
\mathcal{H}_{1} &= \left[ \frac{2h\lm}{1+\lm^2} -\frac{4 h \qq \lm (1-\lm^2)}{(1+\lm^2)^2}\right] \sum_{i}
\left( \Tb_{i}^{\dagger} + \Tb_{i} \right) \,, %\\
\end{align}
%%%%%
\begin{widetext}
\begin{align}
\label{eq:iso_h2}
\mathcal{H}_{2} &= - \frac{J}{(1+\lm^2)^2} \sum_{\langle ij \rangle} \bigg[ (1-\lm^2)^{2} \left( \Tb_{i} \Tb_{j} +
\Tb_{i}^{\dagger} + {\rm h.c.} \right) - 8\lm^{2} \left( \Tb_{i}^{\dagger} \Tb_{i} + \Tb_{j}^{\dagger} \Tb_{j} \right) \bigg] %\nonumber \\
%&~~~~
+ \frac{2 h (1-\lm^2)}{1+\lm^2} \sum_{i} \Tb_{i}^{\dagger} \Tb_{i} \,, \\
%\end{align}
%%%%%%
%\begin{align}
\label{eq:iso_h3}
\mathcal{H}_{3} &= \frac{4 J \lm (1-\lm^2)}{(1+\lm^2)^2} \sum_{\langle ij \rangle} \bigg[ \Tb_{i}^{\dagger} \Tb_{i}^{\dagger} \Tb_{i} + 2 \Tb_{j}^{\dagger} \Tb_{i}^{\dagger} \Tb_{i} + {\rm h.c.} \bigg] %\nonumber \\
- \frac{2 h \lm}{1+\lm^2} \sum_{i}  \bigg[ \Tb_{i}^{\dagger} \Tb_{i}^{\dagger} \Tb_{i} + {\rm h.c.} \bigg] \,, \\
%%%%%%
\label{eq:iso_h4}
\mathcal{H}_4 &= \frac{2 J}{ (1+ \lm^2)^2} \sum_{\langle ij \rangle} \bigg[ (1 - \lm^2)^2
\left( \Tb_{i}^{\dagger} \Tb_{i} \Tb_{i} \Tb_{j} + \Tb_{i}^{\dagger} \Tb_{i}^{\dagger} \Tb_{i} \Tb_{j} + {\rm h.c.} \right)
- 8 \lm^2 \Tb_{i}^{\dagger} \Tb_{j}^{\dagger} \Tb_{i} \Tb_{j} \bigg] \,.
\end{align}
\end{widetext}
Again note that for $\lm=0$ we recover the Hamiltonian in the disordered phase in terms of $T$ operators.

Vanishing of $\mathcal{H}_{1}$ \eqref{eq:iso_h1} gives the leading order $\lm$, denoted by $\lm_0$ and is given in terms of $\qq$ as
\begin{equation}
\label{eq:iso_lam0}
\lm_{0} = \sqrt{\frac{2 \qq - 1}{2 \qq + 1}} \,.
\end{equation}
Using this we will separate the unperturbed piece from the bilinear Hamiltonian \eqref{eq:iso_h2} in the Fourier space, which is given by
\begin{equation}
\label{eq:iso_h2k}
\mathcal{H}_{2 \kk}^{(0)} = \sum_{\kk} \left[ \Ab_{\kk}^{(0)} \Tb_{\kk}^{\dagger} \Tb_{\kk} + \frac{\Bb_{\kk}^{(0)}}{2} \left( \Tb_{\kk} \Tb_{-\kk} + {\rm h.c.} \right) \right]
\end{equation}
with
\begin{equation}
\label{iso_bareAB1}
\Ab_{\kk}^{(0)} = \frac{h}{\qq} + \frac{h(4 \qq^2 -1)}{\qq} + \Bb_{\kk}\,,~~
\Bb_{\kk}^{(0)} = - \frac{\gk h}{2\qq} \,.
\end{equation}
It is then diagonalized using bosonic Bogoliubov transformation
\begin{equation}
\label{iso_bogol}
\Tb_{\vec{k}} =
\ub_{\vec{k}}\tb_{\vec{k}}+\vb_{\vec{k}}\tb^\dagger_{-\vec{k}} \,,
\end{equation}
with the Bogoliubov coefficients having a similar expression to eq. \eqref{eq:isd_bogco} in terms of $\Ab_{\kk}^{(0)}$, $\Bb_{\kk}^{(0)}$, and $\tilde{\omega}_{\kk}$, with
\begin{equation}
\label{eq:iso_om0}
\tilde{\omega}_{\kk} = 2 h \sqrt{4 \qq^2 - \gk} \,.
\end{equation}
The perturbation strategy is then same as in the paramagnetic phase with complications arising from $1/d$ expansion of $\lm$. A detailed account on how to proceed with the diagrammatics in the presence of a $1/d$ expansion to $\lm$ can be found in Ref. \onlinecite{Joshi2015b}. Now we quote the expressions for normal-ordered Hamiltonian in the $\tb$ basis, relevant to order $1/d$. The bilinear piece has three contributions: (i) Unperturbed (diagonal) piece $\mathcal{H}'_{2 a}$, (ii) Bilinear terms, $\mathcal{H}'_{2 b}$, arising from normal ordering of quartic terms, and (iii) Order $1/d$ terms, $\mathcal{H}'_{2 c}$, arising from \eqref{eq:iso_h2} due to $1/d$ corrections to $\lm$. So normal-ordered bilinear Hamiltonian is $\mathcal{H}'_{2} = \mathcal{H}'_{2 a} + \mathcal{H}'_{2 b} + \mathcal{H}'_{2 c}$. We now quote the explicit expressions as follows:
\begin{align}
\label{eq:iso_h2a}
\mathcal{H}'_{2 a} &= \tilde{\omega}_{\kk} \tb^{\dagger} \tb \,, \\
\mathcal{H}'_{2b} &= \sum_{\kk} \bigg[ C_{\kk} \tau_{\kk}^{\dagger} \tau_{\kk}
+ \frac{D_{\kk}}{2} (\tau_{\kk} \tau_{-\kk} + {\rm h.c.} )\bigg] \,, \\
\mathcal{H}'_{2c} &= \sum_{\kk} \bigg[ \left( (\ub_{\kk}^2 + \vb_{\kk}^2) \Ab_{\kk}^{(1)} + 2 \ub_{\kk} \vb_{\kk} \Bb_{\kk}^{(1)} \right) \tau_{\kk}^{\dagger} \tau_{\kk} \nonumber \\
&+ \frac{ (\ub_{\kk}^2 + \vb_{\kk}^2) \Bb_{\kk}^{(1)} + 2 \ub_{\kk} \vb_{\kk} \Ab_{\kk}^{(1)} }{2} (\tau_{\kk} \tau_{-\kk} + {\rm h.c.})\bigg] \,,
\end{align}
where
\begin{widetext}
\begin{align}
\label{eq:iso_c}
C_{\kk} &= 4 \qq h \left(\frac{1-\lm^2}{1+\lm^2}\right)^{2} \big[ (\ub_{\kk}^{2} + \vb_{\kk}^{2}) (\gk R_{1} + 2 \gk R_{2} + 2R_{3}) %\nonumber \\
+ 2 \ub_{\kk} \vb_{\kk} (\gk R_{1} + 2 \gk R_{2} + R_{3}) \big] \nonumber \\
&-\frac{16 \lm^{2} \qq h}{(1+\lm^2)^2} \big[ 2 (\ub_{\kk}^{2} + \vb_{\kk}^{2}) R_2 + 4 \ub_{\kk} \vb_{\kk} \gk R_3 \big] \,, \\
%%%%%%%%%
\label{eq:iso_d}
D_{\kk} &= 4 \qq h \left(\frac{1-\lm^2}{1+\lm^2}\right)^{2} \big[ (\ub_{\kk}^{2} + \vb_{\kk}^{2}) (\gk R_{1} + 2 \gk R_{2} + R_{3}) %\nonumber \\
+ 2 \ub_{\kk} \vb_{\kk} (\gk R_{1} + 2 \gk R_{2} + 2R_{3}) \big] \nonumber \\
&-\frac{32 \lm^{2} \qq h}{(1+\lm^2)^2} \big[ (\ub_{\kk}^{2} + \vb_{\kk}^{2}) \gk R_3 + 2 \ub_{\kk} \vb_{\kk} R_2 \big] \,, \\
%%%%%%%%%%
\label{eq:iso_a1}
\Ab_{\kk}^{(1)} &= -\frac{2 h}{\qq} \left( R_2 + R_3 \right) \left( 1 + \gk \right) \,, ~~~~~~
\Bb_{\kk}^{(1)} = -\frac{2 \gk h}{\qq} \left( R_2 + R_3 \right) \,.
\end{align}
\end{widetext}
The expressions for $R$'s are as follows:
\begin{align}
\label{eq:iso_r1}
R_1 &= \frac{1}{N} \sum_{\vec k} \ub_{\vec k} \vb_{\vec k} ~~~= \frac{1}{256 \qq^4 d} + \mathcal{O}(d^{-2}) \,, \\
%%%%%%%
\label{eq:iso_r2}
R_2 &= \frac{1}{N} \sum_{\vec k} \vb^2_{\vec k} ~~~~~~= \frac{1}{512 \qq^4 d} + \mathcal{O}(d^{-2}) \,, \\
%%%%%%%
\label{eq:iso_r3}
R_3 &= \frac{1}{N} \sum_{\vec k} \gk \ub_{\vec k} \vb_{\vec k} = \frac{1}{32 \qq^2 d} + \mathcal{O}(d^{-2}) \,, \\
%%%%%%%%
\label{eq:iso_r4}
R_4 &= \frac{1}{N} \sum_{\vec k} \gamma_{\vec k} \vb^2_{\vec k} ~~~\,= \mathcal{O}(d^{-2}) \,.
\end{align}

Now, the normal-ordered cubic terms and relevant vertex functions are as follows:
\begin{align}
\label{eq:iso_h3p}
\mathcal{H}'_{3} &= \frac{1}{\sqrt{N}} \sum_{123} \big[ \delta_{1+2+3} \vcao (\tb_{1}^{\dagger} \tb_{2}^{\dagger} \tb_{3}^{\dagger} + \tb_{1} \tb_{2} \tb_{3}) \nonumber \\
&+ \delta_{1+2-3} \vcbo (\tb_{1}^{\dagger} \tb_{2}^{\dagger} \tb_{3} + \tb_{3}^{\dagger} \tb_{2} \tb_{1}) \big] \,,
\end{align}
with
\begin{align}
\label{eq:iso_vca}
\vcao &= (2 J_1 \gamma_{1} - h_1) (\ub_1 \ub_2 \vb_3 + \vb_1 \vb_2 \ub_3) \,, \\
\label{eq:iso_vcb}
\vcbo &= 2 J_1 \big[ \gamma_{1} (\ub_1 \ub_2 \ub_3 + \vb_1 \vb_2 \vb_3 ) \nonumber \\
&+ (\gamma_1 + \gamma_3) (\ub_1 \vb_2 \vb_3 + \vb_1 \ub_2 \ub_3 )  \big] \,,
\end{align}
where
\begin{equation}
\label{eq:iso_h1j1}
h_1 = \frac{2 h \lm}{1 + \lm^2} - J_1 \,, ~~~ \text{and} ~~~
%\label{iso_j1}
J_1 = \frac{4 \qq h \lm (1-\lm^2)}{(1+\lm^2)^2} \,.
\end{equation}
Normal ordering of cubic terms also lead to additional linear terms, which together with $\mathcal{H}_{1}$ \eqref{eq:iso_h1} must vanish. This gives the $1/d$ expansion of $\lm$ as follows:
\begin{equation}
\label{eq:iso_lmexp}
\lm^2 = \frac{2 \qq - 1}{2 \qq + 1} - \frac{1}{64 \qq^3 d} \frac{1 + 16 \qq^2}{(2 \qq +1)^2} \,.
\end{equation}
Again, using the condition that $\lm$ vanishes at the quantum critical point, we obtain the same phase boundary \eqref{eq:isd_pb}.

Lastly, we quote the normal-ordered quartic term and the relevant vertex functions:
\begin{align}
\label{iso_h4p}
\mathcal{H}'_4 &= \frac{1}{N} \sum_{1234}
\big[\delta_{1+2+3+4} \vqao
(\tb^\dagger_{1}\tb^\dagger_{2}\tb^\dagger_{3}\tb^\dagger_{4} +
\tb_{1}\tb_{2}\tb_{3}\tb_{4})   \nonumber \\
&~~~~~~~~~~~
+ \delta_{1+2-3-4} (\vqbo
\tb^\dagger_{1}\tb^\dagger_{2}\tb_{3}\tb_{4}
+\vqco \tb^\dagger_{1}\tb^\dagger_{2}\tb_{3}\tau_{4})
\nonumber \\
&~~~~~~~~~~~+\delta_{1+2+3-4} \vqdo
(\tb^\dagger_{1}\tb^\dagger_{2}\tb^\dagger_{3}\tb_{4} +
\tb^\dagger_{4} \tb_{3}\tb_{2}\tb_{1})\big] \,,
\end{align}
with
\begin{align}
\vqao &= \left(\frac{1-\lm^2}{1+\lm^2}\right)^{2} 2 \qq h \big[ \gamma_{4} \ub_{1} \vb_{2} \vb_{3} \vb_{4} + \gamma_{4} \vb_{1} \ub_{2} \ub_{3} \ub_{4} \nonumber \\
&~~~~~~~~~~~+ (\gamma_{1} + \gamma_{4}) \ub_{1} \ub_{2} \vb_{3} \vb_{4} \big] \nonumber \\
&-\frac{16 \lm^2 \qq h}{(1+\lm^2)^2} \ub_{1} \ub_{2} \vb_{3} \vb_{4} \gamma_{2-4} \,, \\
%%%%%%%%
\vqdo &= \left(\frac{1-\lm^2}{1+\lm^2}\right)^{2} 2 \qq h \big[ (2 \gamma_{3} + \gamma_{4}) \ub_{1} \vb_{2} \vb_{3} \ub_{4}
\nonumber \\
&~~~~~~~~~~~+ (2 \gamma_{3} + \gamma_{4}) \vb_{1} \ub_{2} \ub_{3} \vb_{4} \nonumber \\
&~~~~~~~~~~~+ \gamma_{3} \vb_{1} \vb_{2} \vb_{3} \vb_{4} %\nonumber \\
+ \gamma_{3} \ub_{1} \ub_{2} \ub_{3} \ub_{4}  \nonumber \\
&~~~~~~~~~~~+ (2 \gamma_{1} + \gamma_{3} + \gamma_{4}) \ub_{1} \ub_{2} \vb_{3} \ub_{4} \nonumber \\
&~~~~~~~~~~~+ (\gamma_{1} + 2 \gamma_{3} + \gamma_{4}) \ub_{1} \vb_{2} \vb_{3} \vb_{4} \big]  \nonumber \\
&-\frac{16 \lm^2 \qq h}{(1+\lm^2)^2} \big[  (\gamma_{2-4} + \gamma_{2+3})\ub_{1} \ub_{2} \vb_{3} \ub_{4}  \nonumber \\
&~~~~~~~~~~~~~~~~~~~~~~+ (\gamma_{3-4} + \gamma_{1+3}) \ub_{1} \vb_{2} \vb_{3} \vb_{4}\big] \,.
\end{align}
The diagrammatics to evaluate the dispersion and ground-state energy is same as in the paramagnetic phase, with the additional contribution from the cubic vertices.

%%%%%%%%%%%%%%%%%%%%%%%%%%%%%%%%%%%%%%%%%%%%%%%%%%%%%%%%%%%%%%%%%%%%%%%%%%%%%%%%%%%%%%%

%%%%%%%%%%%%%%%%%%%%%%%%%%%%%%%%%%%%%%%%%%%%%%%%%%%%%%%%%%%%%%%%%%%%%%%%%%%%%%%%%%%%%%%

\end{document}